\documentclass[journal]{IEEEtran}

\usepackage[table, dvipsnames]{xcolor}

\usepackage{amsfonts,amsmath,amssymb,amsthm}
\usepackage{setspace}
\usepackage[noend]{algpseudocode}
\usepackage{xspace}
\usepackage{comment}
\usepackage{graphicx}
\usepackage{multirow}
\usepackage{algorithmicx}

\usepackage{hyperref}
\usepackage[backend=bibtex,bibstyle=ieee,citestyle=numeric-comp,hyperref=true]{biblatex}
\newcommand{\code}{\mathscr C}

\newcommand{\6}{\mathbf}

\algrenewcommand\algorithmicrequire{\textbf{Input:}}
\algrenewcommand\algorithmicensure{\textbf{Output:}}

\usepackage{url}

\usepackage{mathrsfs}

\newtheorem{heuristic}{Heuristic}
\newtheorem{definition}{Definition}

\newtheorem{remark}{Remark}
\newtheorem{theorem}{Theorem}

\newtheorem{proposition}{Proposition}

\usepackage{pgfplots}

\usepackage{calligra}

\newcommand{\hsp}{\hspace{1mm}}

\DeclareFontFamily{U}{BOONDOX-calo}{\skewchar\font=45 }
\DeclareFontShape{U}{BOONDOX-calo}{m}{n}{
  <-> s*[1.05] BOONDOX-r-calo}{}
\DeclareFontShape{U}{BOONDOX-calo}{b}{n}{
  <-> s*[1.05] BOONDOX-b-calo}{}
\DeclareMathAlphabet{\mathcalboondox}{U}{BOONDOX-calo}{m}{n}
\SetMathAlphabet{\mathcalboondox}{bold}{U}{BOONDOX-calo}{b}{n}
\DeclareMathAlphabet{\mathbcalboondox}{U}{BOONDOX-calo}{b}{n}

\usepackage{scalerel}

\usepackage{placeins}

\usepackage{float}

\usepackage[ruled, vlined, linesnumbered]{algorithm2e}

\SetCommentSty{mycommfont}

\usepackage{tikz}

\tikzset{
  symbol/.style={
    draw=none,
    every to/.append style={
      edge node={node [sloped, allow upside down, auto=false]{$#1$}}}
  }
}

\usepackage{tikz}

\tikzset{
  symbol/.style={
    draw=none,
    every to/.append style={
      edge node={node [sloped, allow upside down, auto=false]{$#1$}}}
  }
}

\usepackage{subfig}
\usepackage{amsmath}
\usepackage{mleftright}

\usetikzlibrary{patterns}
\usetikzlibrary{decorations.pathreplacing}

\newcommand{\leb}{\mathsf{LB}}

\usepackage{mathtools}

\usepackage{makecell}

\begin{document}

\title{An Efficient Algorithm to Sample Quantum Low-Density Parity-Check Codes\thanks{
Part of the material in this paper has been submitted for publication at the 2026 IEEE International Symposium on Information Theory (ISIT).
}
}

\author{Paolo Santini\\
Department of Information Engineering, Polytechnic University of Marche, Ancona, Italy \\
\texttt{p.santini@univpm.it}
}

\maketitle

\thispagestyle{plain}
\pagestyle{plain}

\begin{abstract}
In this paper, we present an efficient algorithm to sample random sparse matrices to be used as check matrices for quantum Low-Density Parity-Check (LDPC) codes.
To ease the treatment, we mainly describe our algorithm as a technique to sample a dual-containing binary LDPC code, hence, a sparse matrix $\mathbf H\in\mathbb F_2^{r\times n}$ such that $\mathbf H\mathbf H^\top = \mathbf 0$. 
However, as we show, the algorithm can be easily generalized to sample dual-containing LDPC codes over non binary finite fields as well as more general quantum stabilizer LDPC codes.

While several constructions already exist, all of them are somewhat algebraic as they impose some specific property (e.g., the matrix being quasi-cyclic).
Instead, our algorithm is purely combinatorial as we do not require anything apart from the rows of $\mathbf H$ being sparse enough.
In this sense, we can think of our algorithm as a way to sample sparse, self-orthogonal matrices that are as random as possible.

Our algorithm is conceptually very simple and, as a key ingredient, uses Information Set Decoding (ISD) to sample the rows of $\6H$, one at a time.
The use of ISD is fundamental as, without it, efficient sampling would not be feasible. 
We give a theoretical characterization of our algorithm, determining which ranges of parameters can be sampled as well as the expected computational complexity. 
Numerical simulations and benchmarks confirm the feasibility and efficiency of our approach.
\end{abstract}


\section{Introduction}

For $\mathbb F_q$ a finite field with $q$ elements, we say that a code $\code\subseteq \mathbb F_q^n$ is a $q$-ary LDPC code with length $n$ and dimension $k$ if it admits a parity-check matrix $\6H\in\mathbb F_q^{(n-k)\times n}$ which is sparse, i.e., such that the majority of entries are null.
LDPC codes can be further classified according to several aspects, for instance, whether row and column weights are constant, whether they have some geometrical structure (for instance, quasi cyclicity leading to quasi cyclic LDPC codes), etc.

Due to sparsity, LDPC codes admit efficient encoding and decoding and, for this reason, nowadays find several applications in many communication standards.
Interestingly, LDPC codes are also among the most promising quantum error correcting codes because of good error correction capability and fault tolerance properties. 
For a very well written and up-to-date source about quantum LDPC codes, we refer the reader to \cite{vasic2025quantum}.

For quantum LDPC codes, parity-check matrices must satisfy, in conjunction with sparsity, an extra requirement known as \textit{symplectic product condition}.
This is necessary to guarantee that the subgroup of the Pauli group associated with the code is Abelian, i.e., that its elements commute.
Since this is crucial for this paper, we quickly recall the notions of stabilizer and Calderbank-Shor-Steane (CSS) quantum codes.
Given our scopes (and due to the lack of space), we adopt a linear algebra perspective which allows us to skip the formalism due to quantum physics (and, also, the explanations behind many properties); the interested reader can find all the required explanations in papers from the literature (e.g., \cite{vasic2025quantum,  babar2015fifteen}).
\smallskip

\paragraph*{\textbf{Quantum stabilizer codes}} 
Let $\6H_X, \6H_Z\in\mathbb F_2^{r\times n}$ with full rank and such that\footnote{Even though not strictly necessary, we are using the pedices $X$ and $Z$ as it is common in the literature.
This is due to the relation with Pauli operators which, as we have already stressed, will not be recalled in this paper.}
\begin{equation}
\label{eq:symplectic}
\6H_X\6H_Z^\top+\6H_Z\6H_X^\top = \60.
\end{equation}
The above equality is the already mentioned symplectic product condition.
Then, we can associate these matrices with a stabilizer code of dimension $n-r$, defined by the check matrix $(\6H_X, \6H_Z)\in\mathbb F_2^{r\times 2n}$.
The check matrix can be thought of as the counterpart of parity-check matrices for classical codes: for instance, for a quantum LDPC code, this matrix must be sparse.

By CSS construction, we refer to a special class of stabilizer codes in which two classical codes are combined to get a quantum stabilizer code.
Let $\6H_1\in\mathbb F_2^{r_1\times n}$ and $\6H_2\in\mathbb F_2^{r_2\times n}$ with full ranks $r_1$ and $r_2$, respectively, and such that
\begin{equation}
\label{eq:symplectic_css}
\6H_1\6H_2^\top = \60. \end{equation}
Then, the corresponding CSS code with dimension $n-r_1-r_2$ is the quantum code associated with the check matrix
$\begin{pmatrix}\6H_1 & \60 \\ \60 & \6H_2\end{pmatrix}$.
This construction is just a special case of the previous case, in which $\6H_X$ and $\6H_Z$ have a special (and perhaps convenient) form.
Analogously, \eqref{eq:symplectic_css} follows as a special case of \eqref{eq:symplectic}.

Finally, as a more special case, one can start from two classical codes that are actually the same: in this case, we have $\6H_1 = \6H_2 \coloneqq \6H\in\mathbb F_2^{r\times n}$ and define the check matrix for the CSS code as $\begin{pmatrix}\6H & \60\\\60 & \6H\end{pmatrix}$.
The symplectic product condition turns into
\begin{equation}
\label{eq:symplectic_dual}
\6H\6H^\top = \60,
\end{equation}
which, again, is a specialization of both \eqref{eq:symplectic} and \eqref{eq:symplectic_css}.
In this case, we speak of \textit{dual-containing codes} because, if we indicate with $\code\subseteq \mathbb F_2^n$ the linear code whose parity-check matrix is $\6H$, \eqref{eq:symplectic_dual} implies
$\code^\bot \subseteq \code$.

All of the above theory can be generalized to codes defined over arbitrary finite fields \cite{qudit, qudit2}.
\smallskip

\paragraph*{\textbf{Existing constructions for quantum LDPC codes}}

One may hope to build quantum codes out of classical LDPC codes that guarantee the symplectic product condition.
This may even allow to inherit (with proper caution) the many results about classical LDPC codes such as constructions, distance properties, decoding algorithms, etc.
This is easier said than done!
Indeed, classical LDPC codes typically do satisfy such a condition and, for quantum LDPC codes, ad-hoc constructions are required.

Nowadays, there are several manners to build quantum codes.
For instance, one can start from good classical LDPC codes\footnote{For instance, a good code can be a code with high minimum distance.} and then combine them using the hypergraph-product method to get a CSS LDPC code \cite{tillich2013quantum}.
Concretely, starting from two parity-check matrices $\6H_a \in \mathbb F_2^{r_a\times n_a}$ and $\6H_b\in\mathbb F_2^{r_b\times n_b}$, one obtains the matrices for the stabilizer CSS code as
$$\6H_1 = (\6H_a\otimes \6I_{n_b},\hsp \6I_{r_a}\otimes \6H_b^\top),\quad \6H_2 = (\6I_{n_a}\otimes \6H_{b},\hsp \6H_a^\top \otimes \6I_{r_a}),$$
where $\otimes$ denotes the Kronecker product and $\6I_m$ is the identity of side $m$.

Another popular construction is the so-called \textit{bycycle code} \cite{mackay2004sparse}: given a sparse circulant matrix $\6A \in\mathbb F_2^{m\times m}$, one can get a CSS LDPC code with the dual-containing construction using
$$\6H = (\6A,\hsp\6A^\top).$$
The symplectic product condition \eqref{eq:symplectic_dual} is trivially satisfied, but the resulting quantum code has dimension $O(1)$.
This construction has been later generalized in \cite{generalized_bycicle, generalized_bycicle_2, lin2024quantum} by, e.g., using two circulant matrices and/or matrices with an algebraic regularity different than the cyclic one.
Other constructions are based on quasi-cyclic codes \cite{4557323}, product codes \cite{ostrev2024classical}, dyadic and quasi-dyadic codes \cite{martinez2025codes, baldelli2025quantum}.
Again, we refer the reader to \cite{vasic2025quantum, babar2015fifteen} for excellent surveys of the most prominent constructions for quantum LDPC codes.

All these constructions from the literature exhibits a common feature: they rely on algebraic tools and, arguably, non trivial techniques in order to guarantee both sparsity and the symplectic product condition. 
The resulting codes are  somewhat special and far from being random; moreover, in many cases, their parameters must follow rigid constraints due to the underlying algebraic structure.

This hints at a huge gap between the design of classical and quantum LDPC codes.
Indeed, picking a classical LDPC code is as simple as sampling a matrix with very few ones\footnote{We are obviously aware of the fact that the design of LDPC codes is actually much more convoluted since state-of-the-art solutions consider also relevant quantities such as cycles and trapping sets. What we mean, here, is that a random sparse matrix with proper parameters will yield an LDPC for which a non trivial amount of errors can be corrected efficiently.}.
For quantum codes, instead, we currently have to undergo ad-hoc constructions which, in many cases, limit the range of parameters (e.g., sparsity level, length and dimension).

Very recently, this gap has been narrowed \cite{okada2025random}.
Indeed, in \cite{okada2025random}, the authors start with a structured sparse matrix that satisfies the symplectic product condition (for instance, this is constructed by tiling identity matrices) and then do local modifications to introduce some randomness.
While these codes are arguably less structured than other constructions from the literature, they are still far from being random.
Indeed, the hard part of constructing quantum LDPC codes (that is, sampling a matrix guaranteeing the symplectic product condition) is again fulfilled by sampling a matrix with a special structure.
\smallskip

This state of affairs leads to the following, rather spontaneous question:
\smallskip

\textit{Can random quantum LDPC codes be generated efficiently?}


\subsection{Our contribution}

This paper provides a rather positive answer to the above question.
We show that, under some caveats and some (plausible) heuristics, random quantum LDPC codes can indeed be sampled efficiently, without any compelling constraint on the code parameters (apart from the check matrix being sparse enough).
We primarily describe our techniques speaking of dual-containing codes (as this makes the treatment somewhat easier) but, as we show later on, our ideas can be used to sample more general stabilizer LDPC codes.

Concretely, we propose an algorithm to sample rows of a matrix $\6H\in\mathbb F_2^{r\times n}$, one by one, while guaranteeing both sparsity and self-orthogonality.
The algorithm can be divided into $r$ steps: at step $u$, we sample a sparse vector from the dual of the space generated by the already found $u-1$ vectors $\{\6h_1,\hsp \cdots,\hsp \6h_{u-1}\}$.
This is done by seeing $\6H_u = \begin{pmatrix}\6h_1\\ \vdots \\ \6h_{u-1}\end{pmatrix}$ as the parity-check matrix for a linear code $\code_u\subseteq \mathbb F_2^n$ with dimension $n-(u-1)$ and looking among its codewords to find another sparse vector $\6h_u$.
We then iterate the procedure until we have found all the desired rows.
By design, this guarantees that $\6H\6H^\top = \60$.
\medskip

Interestingly, we take a fundamental inspiration from the realm of post-quantum cryptography.
Indeed, the sampling procedure is adapted from the one in \cite{albrecht2025hollow}, where the authors describe how to sample a random self-orthogonal code over $\mathbb F_q$.
We tweak their algorithm by requiring that, at each step, $\6h_u\in \code_u$ is sparse.
To efficiently find such a codeword, we rely on Information Set Decoding (ISD), which is a family of algorithms designed to find the minimum distance of arbitrary linear codes and is the leading attack for many code-based cryptosystems (see e.g. \cite{prange1962, lee1988observation, stern1989method, becker2012decoding, may2011decoding, bernstein2008attacking} for ISD over the binary finite field and \cite{peters2010information, carrier2024projective, elbro2025can, elbro2025decoupling} for arbitrary fields).

For the sake of simplicity, we study our algorithm assuming that all rows have the same weight $v$, but with little effort the whole analysis can be generalized to other weight distributions.
We first derive the conditions that guarantee that, at each step, a codeword of the desired weight can be found.
To this end, we focus on codes having constant code rate $R\in [0 \hsp;\hsp 1]$.
For such parameters, according to the Gilbert Varshamov (GV) distance, random codes have minimum distance which is linear in $n$ with overwhelming probability. 
To sample matrices that are adequately sparser than random matrices, we focus our analysis on the case $v = o(n)$.
We show that, asymptotically, $v<\frac{\ln(2)}{1-R}\cdot \log_2(n)$ grants that, at each step of the algorithm, codewords of weight-$v$ exist with high probability.

To confirm the feasibility and efficiency of our algorithm, we have developed a Sagemath proof of concept implementation which can be found at the following link\footnote{We make publicly available a proof of concept implementation of our sampling algorithm, as well as other scripts that we have used to produce the results in the paper.}: 

\begin{center}
\url{https://github.com/psantini1992/DualContaining_LDPC}
\end{center}

In Table \ref{tab:intro} we report some examples of the codes we have been able to generate, together with the average time, measured on a standard laptop (more details can be found in Section \ref{sec:results}).
To compare with random codes, in the table we report also the GV distance for random codes with the same parameters $n$ and $r$.

We remark that there is wide room for improving timings since our implementation is only a proof-of-concept and, moreover, we have used one of the most basic versions of ISD. 
Since most of the time is spent exactly in the enumeration subroutine of ISD, timings can be greatly reduced by switching to an optimized implementation and/or to a more advanced ISD (and, as we argue next, to an ISD that is specifically tailored for LDPC codes).

\begin{table}[t]
\centering
\begin{tabular}{|c|c|c|c|c|}
\hline
$n$                  & $r$                  & GV distance           & $v$ & Avg. Time (seconds)\\ \hline\hline
\multirow{5}{*}{250} & \multirow{5}{*}{80} & \multirow{5}{*}{$16$} & $6$ & $2.13$\\  
&         && $8$ & $2.31$\\ 
&         && $10$ & $3.62$\\ 
&         && $12$ & $22.84$\\ 
&         && $14$ & $412.19$\\ 
\hline
\multirow{3}{*}{500} & \multirow{3}{*}{200} & \multirow{3}{*}{$41$} & $6$ & $22.77$ \\           
& &  & $8$ &  $25.40$ \\ 
          &                      &                       & $10$ &  $164.05$     \\
\hline
\multirow{3}{*}{1000} & \multirow{3}{*}{400} & \multirow{3}{*}{$81$} & $6$ & $178.99$ \\  
                     &                      &                       & $8$ & $235.01$ 
                     \\ 
                      
                     &                      &                       & $10$ & $405.06$        \\\hline
\end{tabular}
\caption{Average time required to sample self-orthogonal codes using a Sagemath proof-of-concept implementation of our algorithm. The code has been tested on a standard laptop. For each parameters set, timings have been averaged over 10 runs.}
\label{tab:intro}
\end{table}

In Section \ref{sec:generalizations} we discuss simple tweaks to sample more general types of codes.
For instance, as we have already said, one may choose a less rigid weight distribution for the rows of $\6H$.
The algorithm can be also tweaked to sample self-orthogonal codes over non binary finite fields.
This comes with a very mild overhead, since ISD shall now work over a $q$-ary field and, moreover, the sampling of each row shall be repeated on average for $q$ times\footnote{If the finite field has characteristic 2, i.e., $q = 2^\ell$ for some integer $\ell$, the latter overhead actually does not apply.}. 
Dulcis in fundo, we show also how to sample more general quantum LDPC codes, even stabilizer codes that do not follow the CSS construction.


\subsection{Technical overview}

In this section we discuss about some relevant technical aspects of our analysis.
In particular, we want to clarify what we are able to achieve, what we can probably achieve (but are not able to prove) and what remains as an open question.
Moreover, we also want to emphasize limitations of this work and, eventually, state what we see as interesting follow up works.

\paragraph*{\textbf{Achievable values of $v$}}

As we have already said, we shall choose $v$ so that, at every step, a solution with the desired weight is guaranteed to exist.
We do this by modeling $\6H_u$ as a sample from the distribution $\mathcal H_{n,u,v}$ that returns uniformly random $u\times n$ matrices with row weight $v$.
Then, we treat $\6H_u$ as the parity-check matrix of a code $\code_u$ and derive its Expected Weight Distribution (EWD).
This analysis is non trivial and, perhaps, provides a contribution of independent interest.

We show that when $v$ is small enough, then $\code_u$ has codewords of small weight.
In particular, we prove that, asymptotically, $\code_u$ has on average codewords of weight $o(n)$ only if $v<\frac{\ln(2)}{1-R}\cdot \log_2(n)$.

This analysis allows us to define which values of $v$ we shall use.
Namely, if for any $u\in\{1,\hsp\cdots,\hsp r-1\}$, the value of $m_v$ (which is the expected amount of weight-$v$ codewords) is greater than 1, then our algorithm shall succeed with large probability.
If this condition is not fulfilled, then our algorithm is likely going to get stuck because, in some intermediate step, it may not be able to find codewords of the desired weight.
\medskip


\paragraph*{\textbf{Can we claim polynomial time?}}

A well known result states that, for a sublinear weight $w$, ISD runs in time $\frac{1}{m_w}\cdot 2^{w\cdot\log_2(1-R)\cdot \big(1+o(1)\big)}$\cite{canto2016analysis}.
In our case, we use ISD to search for codewords of weight $v = o(n)$, which already yields sub-exponential time.
In particular, whenever $v = O\big(\log(n)\big)$, this yields polynomial time.
Moreover, when $m_v$ is exponentially large, we may achieve polynomial time as well since, in these cases, ISD boils down to enumerating vectors with weight $p$, with $p$ being a small constant (say, $p = 2$ or $p = 3$).
Whenever this is true, our algorithm runs in worst case time $O\big(n^2\cdot \binom{n}{p}\big) = O\big(n^{p+2}\big)$.
We discuss about this in Section \ref{sec:isd}; as we shall see in Section \ref{sec:results}, numerical simulations apparently confirm this claim for many cases.

All in all, claiming polynomial time for our algorithm looks rather tempting.
However, we refrain from making such a claim because, as we shall see, this would need to rely on some heuristics which require some caution.

Looking ahead, the most suspicious heuristics are those we use to model the behavior of ISD.
Even though we rely on the same heuristics which are used in all the many papers about ISD, there is a main difference: all of these papers target random codes, while the codes we are concerned with in this paper are LDPC.
In such a case, heuristics shall be taken with a grain of salt and, eventually, adjusted to take sparsity into account.
However, we are not aware of any paper studying ISD for LDPC codes.
We see this as an interesting research direction and plan to study this matter in follow up works.
\medskip

\paragraph*{\textbf{Are our codes truly random? Are they good codes?}}

Remember that we started with the goal of generating random self-orthogonal LDPC codes.
While we are not able to formally prove this (i.e., we are not able to prove that our matrices are uniformly random samples from the set of all self-orthogonal matrices with row weight $v$), we do not see any evident argument to refute this thesis.
Indeed, using ISD, at the $u$-th step we sample at random from the set of all weight-$v$ vectors that are orthogonal to each of the already found vectors $\6h_1,\hsp\cdots,\hsp\6h_{u-1}$.
Thus, apart from self-orthogonality, we do not impose any geometrical nor algebraic structure.

Obviously, this means also that we do not guarantee certain conditions which would make our matrices good parity-check matrices for LDPC codes.
For instance, we do not guarantee that null columns do not exist (but, as we discuss in Section \ref{sec:isd}, getting rid of such columns is easy).
Remarkably, as we show in Section \ref{sec:results}, the distribution of columns weight can be well predicted by studying our matrices as samples from $\mathcal H_{n,r,v}$. 
In any case, by puncturing or shortening, one can get rid of columns having no set entry or, more generally, of columns with too low Hamming weight. 
This will preserve self-orthogonality but will modify a bit the code parameters; we discuss this in Section \ref{sec:isd}.
Analogously, we do not care about cycles and girth, which are fundamental quantities for obtaining good LDPC codes.
We believe such extra conditions can be somehow embedded into our algorithm and leave exploring these aspects for future works.

\section{Notation and background}
In this section we settle the notation we are going to use and recall the necessary background notions about linear codes and ISD.

\subsection{Mathematical notation}

For $q$ a prime power, we indicate with $\mathbb F_q$ a finite field with $q$ elements.
For the majority of the paper, we consider the binary finite field, i.e., the case of $q = 2$.
We use bold capital letters to denote matrices and small capital letters to denote vectors, e.g., $\6A$ is a matrix and $\6a$ is a vector.
The null matrix is indicated as $\60$ (its dimensions will always be clear from the context). 
For a vector $\6a$, we use $\mathrm{wt}(\6a)$ to denote its Hamming weight, that is, the number of non null coordinates.

Given a list of $u$ length-$n$ vectors $\6a_1,\hsp \cdots,\hsp \6a_u$, we denote by $\mathrm{Span}(\6a_1,\hsp \cdots,\hsp  \6a_u)$ the space of all linear combinations of such vectors, and as
$\mathrm{Ker}(\6a_1,\hsp \cdots,\hsp  \6a_u)$ the space of all vectors which are orthogonal to each $\6a_i$.
For a matrix $\6A$, $\mathrm{Ker}(\6A)$ denotes its right kernel, i.e., the space of vectors $\6c$ such that $\6A\6c^\top = \60$.
The Hamming sphere with radius $w$ is the set of all vectors with length $n$ and weight $w$ and is indicated as $\mathcal E_{w}$. 

We will use $S_n$ to indicate the group of length-$n$ permutations.
We see its elements as matrices and, for a matrix $\6P$, write $\6P\in S_n$ if it is a permutation matrix (i.e., all rows and columns have weight 1).

For a distribution $\mathcal D$, we indicate the sampling of $x$ according to $\mathcal D$ as $x\gets \mathcal D$.
The probability that $x$ is returned from $\mathcal D$ is indicated as $\mathcal D(x)$.


\subsection{Asymptotics}

We make an extensive use of Landau's notation.
For functions $f(n)$ and $g(n)$, with $n$ positive, we write $f(n)\sim g(n)$ if $\lim_{n\rightarrow \infty} \frac{f(n)}{g(n)} = 1$.
For the binomial coefficient, we will frequently rely on the following asymptotic expansions:
\begin{itemize}
\item[- ] if $w$ is sublinear in $n$, i.e., $w = o(n)$, then asymptotically
$$\log_2\binom{n}{w} = w\cdot\log_2\left(\frac{n}{w}\right)\cdot\big(1+o(1)\big);$$
\item[- ] if $w$ is linear in $n$, i.e., $w = \Theta(n)$, let $\omega\in\mathbb R_{\geq 0}$ such that $\omega/n\sim \omega$.
Then, asymptotically 
$$\log_2\binom{n}{w} = n\cdot h(\omega)\cdot\big(1+o(1)\big),$$
where $h(x) = -x\log_2(x)-(1-x)\log_2(1-x)$ is the binary entropy function.
\end{itemize}

\subsection{Linear codes}

Given a finite field $\mathbb F_q$ with $q$ elements, a $q$-ary linear code with length $n$ and dimension $k\leq n$ is a $k$-dimensional subspace $\mathscr C\subseteq \mathbb F_q^n$. 
The parameter $r = n-k$ is called redundancy, while the ratio $R = k/n$ is referred to as code rate.

Every linear code can be represented using a generator matrix, that is, a full rank matrix $\6G\in\mathbb F_q^{k\times n}$ such that 
$$\code = \left\{\6u\6G\mid \6u\in\mathbb F_q^k\right\}.$$
Another representation can be obtained using the parity-check matrix, which is a full rank matrix $\6H\in\mathbb F_q^{(n-k)\times n}$ such that$$\code = \left\{\6c\in\mathbb F_2^n\mid \6H\6c^\top = \60\right\}.$$
For any non singular $\6S\in\mathbb F_q^{k\times k}$, we get another generator matrix for $\code$ as $\6S\cdot \6G$.
Analogously, for any non singular $\6S\in\mathbb F_q^{(n-k)\times (n-k)}$, the matrices $\6H$ and $\6S\cdot \6H$ are parity-check matrices for the same code.

The weight distribution of a code $\code$ is given by $\{m^{\mathscr{C}}_0, m^{\mathscr{C}}_1,\ldots m^{\mathscr{C}}_n\}$, where $m^{\code}_w$ counts the number of weight-$w$ codewords in $\code$.

\paragraph*{Distributions for codes}
Let $\mathcal D$ be a distribution for linear codes with length $n$.
If $\code\subseteq \mathbb F_q^n$ is a linear code that is output by $\mathcal D$ with some non null probability, i.e., $\mathcal D(\code)\neq 0$, then we write $\code \in \mathcal D$.
For a distribution $\mathcal D$, we define the \textit{Expected Weight Distribution (EWD)} as 
$\{m^{\mathcal D}_0,\hsp m^{\mathcal D}_1,\hsp \cdots,\hsp m^{\mathcal D}_n\}$, where each value $m_w^{\mathcal D}$ is obtained by averaging over all codes which can be sampled from $\code$, that is,
$$m^{\mathcal D}_w = \sum_{\code \in \mathcal D}m^{\code}_w\cdot  \mathcal D(\code).$$
In this paper, we will mostly describe distributions for codes that, in practice, return parity-check matrices.
Accordingly, we modify the EWD as 
$$m^{\mathcal D}_w = \sum_{\6H \in \mathcal D}m^{\6H}_w\cdot  \mathcal D(\6H),$$
where the term $\6m^{\6H}_w$ is the $w$-th coefficient of the weight distribution of the code whose parity-check matrix is $\6H$.

Given a distribution $\mathcal D$ for codes, we denote as Expected Minimum Distance (EMD) the minimum value of $w$ for which $m^{\mathcal D}_w\geq 1$.

\subsection{Random binary codes}

By \textit{random binary code}, we mean a code whose parity-check matrix is sampled uniformly at random from $\mathbb F_2$.

As it is well known, the EWD for this distribution can be derived as follows.
\begin{theorem}
For parameters $r,n\in\mathbb N$, let $\mathcal U_{n,r}$ be the uniform distribution over $\mathbb F_2^{r\times n}$.
Then, the $w$-th coefficient of the EWD is 
$$m^{\mathcal U_{n,r}}_w = \binom{n}{w}2^{-r}.$$
\end{theorem}
\begin{remark}
We do not prove this theorem now as, later on, we will provide a more general proof working for a broader distribution.
\end{remark}
Since the values of $n$ and $r$ will always be clear from the context, we avoid overloading the notation and simply indicate the coefficients of the EWD for random codes as $m^{\mathrm{rnd}}_w$.

We finally recall the notion of Gilbert Varshamov (GV) distance.
\begin{theorem}
For parameters $r,n\in\mathbb N$, we define the Gilbert Varshamov (GV) distance as
$$d = \min_{w\in\mathbb N}\left\{m^{\mathrm{rnd}}_w\geq 1\right\}.$$
For $r = (1-R)n$ with constant $R$, we have 
$$d = \delta n\cdot \big(1+o(1)\big),$$
where $\delta = h^{-1}(1-R)$.
\end{theorem}
Observe that the GV distance is the EMD for the distribution of random codes.

\subsection{Information Set Decoding}

By ISD, we refer to a family of algorithms that, on input an arbitrary linear code and a target weight $w$, returns (if it exists) a codeword of weight $w$.
ISD is a randomized algorithm, in the sense that whenever there are multiple codewords with the same weight, then ISD returns one of these codewords at random.

In this section we recall the functioning of the algorithm proposed by Lee and Brickell \cite{lee1988observation}, which we refer to as $\leb$. 
The expert reader may wonder why we chose to rely on such an algorithm instead of more advanced variants, such as \cite{stern1989method} and \cite{becker2012decoding}.
As we shall see later on, despite its simplicity, $\leb$ is already efficient enough for our purposes.
Moreover, the sparsity of the parity-check matrices we are interested in may makes the analysis of ISD a bit cumbersome (this will be discussed in Sections \ref{sec:isd} and \ref{sec:results}).
For these reasons, in this work we have focused on one of the simplest ISD algorithms and plan to investigate advances algorithms in future works.

Algorithm \ref{alg:isd_general} shows how $\leb$ works.
The operations can be grouped into two main steps:
\begin{itemize}
\item[-] \textit{Permutation and Gaussian Elimination}:
First, sample a random $n  \times n$ permutation $\6P \in S_n$ and then do Gaussian elimination on $\6H\6P$, in order to obtain 
$\6H' = (\6A,\hsp \6I_{n-k})
$, with $\6A\in\mathbb F_q^{k\times (n-k)}$.
Observe that this is not possible whenever the leftmost $n-k$ columns of $\6H\6P$ are linearly dependent; in such a case, start again by sampling a new permutation.
\item[-] \textit{Enumeration and candidates checking}: 
Let $(\6c',\hsp \6c'')$, with $\6c'$ of length $k$ and $\6c''$ with length $n-k$, be a codeword.
Then, it must be
$$\6c'' = -\6A\6c'^\top.$$
The algorithm enumerates all vectors $\6c'$ with weight $p$ and, for each of them, computes $\6c''$ as above.
When $\6c''$ has weight $w-p$, then a codeword of the desired weight is found and the algorithm returns $\6c = (\6c',\hsp\6c'')\6P^{-1}$.
\end{itemize}
\begin{algorithm}[ht]
\KwData{parameter $p\in\mathbb N$, $1\leq p\leq k$}
\KwIn{$\6H\in\mathbb F_q^{(n-k)\times n}$, $w\in\mathbb N$}
\KwOut{failure $\bot$, or vector $\6c\in\mathbb F_q^n$ such that $\6H\6c^\top = \60$ and $\mathrm{wt}(\6c) =w$}
\vspace{0.5em}

\tcp{Permutation and Gaussian elimination}
Sample $\6P$ from $S_n$\;
Set $\6H\cdot \6P = (\6H_1,\hsp \6H_2)$\;
If $\mathrm{Rank}(\6H_2) = n-k$, set $\6A = \6H_2^{-1}\cdot \6H_1$, else restart from instruction 1.

\vspace{0.5em}
\tcp{Enumerate weight-$p$ patterns}
\For{$\6c'\in \mathbb F_q^{k}$, $\mathrm{wt}(\6c') = p$}{
Compute $\6c'' = -\6A\6c'^\top$\;
\If{$\mathrm{wt}(\6c'')= w-p$}{
Set $\6c = (\6c', \hsp \6c'')\cdot \6P^{-1}$\;
\Return $\6c$\;
}
}
\Return $\bot$\;

\caption{$\leb$}
\label{alg:isd_general}
\end{algorithm}
\medskip

We now recall how the complexity of the algorithm is estimated.
We first introduce two heuristics which are instrumental in assessing the analysis of ISD.
\begin{heuristic}\label{heu:info_set}
For a linear code over $\mathbb F_q$, with length $n$ and dimension $k$, let $\6H\in\mathbb F_q^{(n-k)\times n}$ be a parity-check matrix.
Then, for each set $J\subseteq \{1,\hsp \ldots,\hsp n\}$ of size $n-k$, the probability that the columns indexed by $J$ form a non singular matrix is $\gamma_q = \prod_{i = 1}^{n-k-1}1-q^{-i}$.
\end{heuristic}
\begin{heuristic}\label{heu:codewords_w}
For a linear code over $\mathbb F_q$, with length $n$ and dimension $k$, all codewords with weight $w$ behave as independent and uncorrelated samples from the Hamming sphere with radius $w$, over $\mathbb F_q$.
\end{heuristic}
Observe that the first heuristic, in practice, corresponds to assuming that submatrices of a parity-check matrix behave as random.
Indeed, $\gamma_q$ is the probability that a uniformly random $(n-k)\times (n-k)$ over $\mathbb F_q$ matrix is non singular.
The second heuristic is useful when studying how existence of multiple codewords affects the success probability of one iteration of ISD.

Below, we recall how the time complexity of $\leb$ can be estimated.
\begin{proposition}\label{prop:lee}
We consider one call to $\leb$, with parameter $p$, on input $\6H\in\mathbb F_2^{(n-k)\times n}$ and $w\in\mathbb N$.
Under heuristics \ref{heu:info_set} and \ref{heu:codewords_w}, the success probability of the algorithm is 
$$P_{\leb}\left(n,k,w, m_w^{\6H}\right) = 1-\left(1-\binom{k}{p}\binom{n-k}{w-p} / \binom{n}{w}\right)^{m_w^{\6H}}.$$
The time complexity, measured as amount of elementary operations over $\mathbb F_2$, can be estimated as 
$$T_{\leb}(n,k) = \frac{1}{\gamma_2}\cdot (n-k)^2(n+k)+n\cdot \binom{k}{p}.$$
When $m_w^{\code}\geq 1$, repeatedly calling $\leb$ until a weight-$w$ codeword is found, the average time complexity to find a codeword with weight $w$ is $T_{\leb}(n,k)/P_{\leb}\left(n,k,w, m_w^{\6H}\right)$.
\end{proposition}
\begin{IEEEproof}
The cost of one iteration is obtained by estimating as $(n-k)^2(n+k)$ the cost of Gaussian elimination, and as $n\binom{k}{p}(q-1)^p$ the cost of testing all weight-$p$ patterns.
Observe that, under heuristic \ref{heu:info_set}, we estimate as $\gamma_q$ the probability that $\6H_2$ has full rank.

For what concerns the success probability, we consider that for each codeword of weight $w$, $\frac{\binom{k}{p}\binom{n-k}{w-p}}{\binom{n}{w}}$ is the probability that the chosen permutation correctly partitions the weight.
Then, we need to consider the probability that the permutation is valid for at least one codeword.
Under heuristic \ref{heu:codewords_w}, the probability is the same for all codewords of weight $w$.
Thus, the probability we are seeking is the complementary of the probability that the permutation is not valid for all codewords.
\end{IEEEproof}
\begin{remark}
As a simple but precise approximation for the success probability, we have
$$P_{\leb}(n,k,w,m_w^{\code})\approx \min\left\{1\hsp ; \hsp m_w^{\code}\cdot \frac{\binom{k}{p}\binom{n-k}{w-p}}{\binom{n}{w}}\right\}.$$
Notice that when $m_w^{\code} = 0$, the success probability becomes 0.
\end{remark}

\section{Constant Row Weight Parity-Checks and Expected Weight Distribution}
\label{sec:distributions}
In this section we consider only binary codes.
We first define the distribution for codes that we will thoroughly study.
\begin{definition}
For parameters $r,n\in\mathbb N$ and $v\in\mathbb N$, with $r,v\leq n$, we define $\mathcal H_{n, r, v}$ as the distribution that returns $\6H\in\mathbb F_2^{r\times n}$ where each row is sampled uniformly at random from $\mathcal E_v$.
\end{definition}
We now derive the EWD for this distribution.
\begin{theorem}\label{the:ewd_distrib}
The EWD of the distribution $\mathcal H_{n,r,v}$ is given by 
$$m_w = \binom{n}{w}\cdot \rho_w^r,$$
where $$\rho_w = \sum_{\begin{smallmatrix}0\leq i \leq \min\{w\hspace{0.5mm} ; \hspace{0.5mm} v\}\\ \text{$i$ even}\end{smallmatrix}}\frac{\binom{v}{i}\binom{n-v}{w-i}}{\binom{n}{w}}.$$
\end{theorem}
\begin{IEEEproof}
First, we observe that the $w$-th coefficient of the EWD is computed as
\begin{align*}
m_w &\nonumber = \sum_{\6H\in\mathbb F_2^{r\times n}}\left|\mathcal E_{w}\cap \mathrm{Ker}(\6H)\right|\cdot \mathcal H_{n,r,v}(\6H)\\\nonumber
& = \sum_{\6H\in\mathbb F_2^{r\times n}}\sum_{\6c\in \mathcal E_{w}}f(\6c, \6H)\cdot \mathcal H_{n,r,v}(\6H),
\end{align*}
where $f(\6c, \6H) = 1$ if $\6c\in\mathrm{Ker}(\6H)$ and $0$ otherwise.
We can swap the two sums in the above equation and get
\begin{align*}
m_w &\nonumber  = \sum_{\6c\in \mathcal E_{w}}\sum_{\6H\in\mathbb F_2^{r\times n}}f(\6c, \6H)\cdot \mathcal H_{n,r,v}(\6H)\\\nonumber
& = \sum_{\6c\in \mathcal E_{w}}\underbrace{\sum_{\begin{smallmatrix}\6H\in\mathcal H_{n,r,v}\\\6H\6c^\top = \60\end{smallmatrix}}\mathcal H_{n,r,v}(\6H)}_{\mathrm{Pr}\left[\6H\6c^\top = \60\mid \6H\gets \mathcal H_{n,r,v}\right]}.
\end{align*}
We have $\6H\6c^\top = \60$ if and only if $\6h_i\6c^\top = 0$ for each $i$, where $\6h_i$ denotes the $i$-th row of $\6H$.
For each $i$, we model this as a hypergeometric distribution where we take $w$ samples from a population made of $v$ success and $n-v$ failures.
Hence, the probability that $\6h_i\6c^\top = 0$ is
$$\rho_w = \sum_{\begin{smallmatrix}0\leq i \leq \min\{w\hspace{0.5mm} ; \hspace{0.5mm} v\}\\ \text{$i$ even}\end{smallmatrix}}\frac{\binom{v}{i}\binom{n-v}{w-i}}{\binom{n}{w}}.$$
Since all rows of $\6H$ are sampled independently, we raise it to the $r$-th power and finally obtain the probability in the thesis.
\end{IEEEproof}

\begin{figure}[h]
\resizebox{\columnwidth}{!}{
\pgfplotsset{every tick label/.append style={font=\scriptsize}}

\begin{tikzpicture}
    \begin{axis}[
        xlabel={$w$},
        ylabel={$\log_{2}(m_w)$},
        xmin=0, xmax=220,
        ymin=-200, ymax=200,
        xtick={1, 50, 100, 150, 200},
        ytick={-200, -100, 0, 100, 200},
        legend pos = south east,
        legend style={font=\scriptsize},
        xmajorgrids=true,
        ymajorgrids=true,
        height  = 5.5cm,
        major grid style = {gray!20!white},
        width = 1\columnwidth,
        legend columns=1
    ]

\pgfplotstableread{Data/n_1000_r_500_v_5.txt}{\vnine}
\addplot [RoyalBlue, thick] table[x=X, y=Y] {\vnine};
\addlegendentry{$v = 5$};

\pgfplotstableread{Data/n_1000_r_500_v_9.txt}{\vnine}
\addplot [Bittersweet, thick] table[x=X, y=Y] {\vnine};
\addlegendentry{$v = 9$};

\pgfplotstableread{Data/n_1000_r_500_v_13.txt}{\vnine}
\addplot [PineGreen, thick] table[x=X, y=Y] {\vnine};
\addlegendentry{$v = 13$};

\pgfplotstableread{Data/n_1000_r_500_v_17.txt}{\vnine}
\addplot [BurntOrange, thick] table[x=X, y=Y] {\vnine};
\addlegendentry{$v = 17$};

\pgfplotstableread{Data/n_1000_r_500_rnd.txt}{\vnine}
\addplot [Black, dashed, thick] table[x=X, y=Y] {\vnine};


\addplot [gray!50!white, very thick] coordinates{
(0,0)
(220, 0)
};

\pgfplotstableread{Data/n_1000_r_500_v_5.txt}{\vnine}
\addplot [RoyalBlue, thick] table[x=X, y=Y] {\vnine};

\pgfplotstableread{Data/n_1000_r_500_v_9.txt}{\vnine}
\addplot [Bittersweet, thick] table[x=X, y=Y] {\vnine};

\pgfplotstableread{Data/n_1000_r_500_v_13.txt}{\vnine}
\addplot [PineGreen, thick] table[x=X, y=Y] {\vnine};

\pgfplotstableread{Data/n_1000_r_500_v_17.txt}{\vnine}
\addplot [BurntOrange, thick] table[x=X, y=Y] {\vnine};

\end{axis}

\end{tikzpicture}
}
\caption{Initial coefficients for the EWD for codes from $\mathcal H_{n,r,v}$, for $n = 1000$, $r  = 500$ and several values of $v$. The EWD for random codes with the same parameters is plotted as a dashed black curve.}
\label{fig:ewd_toy}
\end{figure}
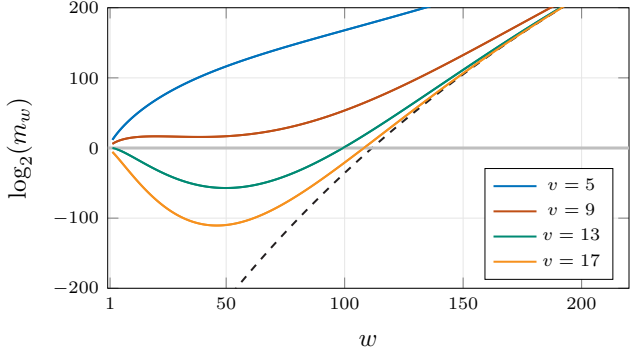

In Figure \ref{fig:ewd_toy} we show some examples of the EWD for $n = 1000$, $r = 500$ and several values of $v$, and compare them with the EWD of random codes with the same parameters.
As we can see, there is a difference between the distribution for codes from $\mathcal H_{n,r,v}$ and that for random codes.
In particular, when $v$ gets smaller, the weight distribution is more biased towards 0.
When $v$ grows, instead, the EWD becomes more similar to that for random codes.
In the next section, we analyze the asymptotic behavior of the derived EWD and give also a justification for this fact.

\subsection{Asymptotic behavior}

We now formalize the intuition that, for small values of $v$, codes sampled from $\mathcal H_{n,r,v}$ have an EWD which is significantly different from that of random codes.
In particular, we show that under certain conditions, there exists values of $w$ which are $o(n)$ and such that $m_w\geq 1$.

Using a well known classical result, when $w = o(n)$, the hypergeometric distribution used for the proof of Theorem \ref{the:ewd_distrib} asymptotically converges to a Binomial distribution with success probability per trial given by $v/n$\footnote{Roughly speaking, sampling with and without replacement does not make a relevant difference for large $n$}. 
We then rewrite the probability $\rho_w$ from Theorem \ref{the:ewd_distrib}.
\begin{theorem}
Let $w = o(n)$.
Then, 
$$\rho_w \sim \frac{1}{2}\cdot \big(1+\varepsilon(n, v, w)\big),$$
where $\varepsilon(n, v, w) = \left(1-\frac{2v}{n}\right)^w$.
\end{theorem}
\begin{IEEEproof}
The proof is based on the observation that, asymptotically, the probability corresponds to the probability that $w$ independent samples of a Bernoulli distribution with parameter $v/n$ sum to 0.
Using the piling-up lemma, this probability corresponds to
$$\frac{1}{2}\cdot\Bigg(1+\underbrace{\left(1-\frac{2v}{n}\right)^w}_{\varepsilon(n, v, w)}\Bigg).$$
\end{IEEEproof}
Recalling now the expression of $m_w$ from Theorem \ref{the:ewd_distrib}, we rewrite it as
\begin{equation}
\label{eq:bew_m_w}
m_w \sim \binom{n}{w}\cdot 2^{-r}\cdot \Big(1+ \varepsilon(n, v, w)\Big)^{r}.
\end{equation}
We will frequently use the logarithm of this quantity (because this makes asymptotics somewhat easier to handle), and we will also consider the case of $r = (1-R)n$ for some constant $R\in [0\hsp;\hsp 1]$.
In such a case, we have that $\log_2(m_w)$ can be expressed as
\begin{equation}
\label{eq:starting_point}
\log_2\binom{n}{w}-n\cdot (1-R)\cdot \Big(1-\log_2\big(1+\varepsilon(n,v,w)\big)\Big).
\end{equation}

Observe that the only difference with respect to the EWD for random codes is in the term $\log_2\Big(1+ \varepsilon(n, v, w)\Big)$.
Whenever $\varepsilon(n, v, w)$ goes to 0, then the EWD for $\mathcal H_{n,r,v}$ converges to the EWD for random codes.
For instance, whenever $v$ is too large, then this is exactly what happens.
We give a proof of this in the following theorem; observe that, in the proofs in this section, we will frequently use the fact that, in the regime $w = o(n)$, we have $\log_2\binom{n}{w} = w\cdot \log_2(n/w)\cdot \big(1+o(1)\big)$.
\begin{theorem}
Let $r = (1-R)n$ for constant $R\in [0\hsp ; \hsp 1]$.
If $v = \nu n$ for some $\nu\in [0 \hsp ; \hsp 1/2]$, then the EMD for the distribution $\mathcal H_{n,r,w}$ approaches that for random codes.
\end{theorem}
\begin{IEEEproof}
In this case, we have $\varepsilon(n, v, w) = (1-2\nu)^w$.
If $w$ grows with $n$, this term goes to $0$, hence $\varepsilon(n, v, w) = o(1)$ for growing $n$.

Observe that $\varepsilon(n, v, w)$ does not go to 0 only when $w$ is constant.
However, in this regime we have $m_w < 1$.
Indeed, $\log_2\binom{n}{w} = O \big(\log_2(n)\big)$, hence asymptotically $\log_2(m_w)$ grows as
$$\log_2(n)-(1-R)\cdot n \cdot \Big(1-\log_2\big(1+(1-2\nu)^w\big)\Big).$$
For growing $n$, the first term grows much slower than the second one which is always negative, hence $m_w\geq 1$ whenever $w$ is constant.
\end{IEEEproof}
We now prove that, instead, we can have an EMD which is sublinear in $n$ whenever $v = o(n)$ and other conditions are fulfilled.
\begin{theorem}\label{the:final_theorem}
Let $r = (1-R)n$ for constant $R\in [0\hsp ; \hsp 1]$.
If $v = o(n)$, then for $w = o(n)$ we can have $m_w\geq 1$ only if i) $vw = o(n)$, and ii) $w\leq n\cdot 2^{\frac{-(1-R)v}{\ln(2)}}$.
\end{theorem}
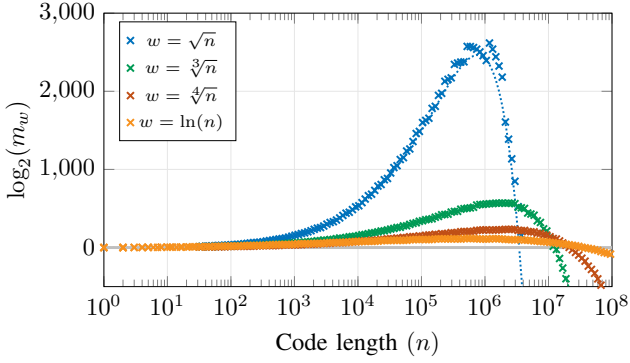
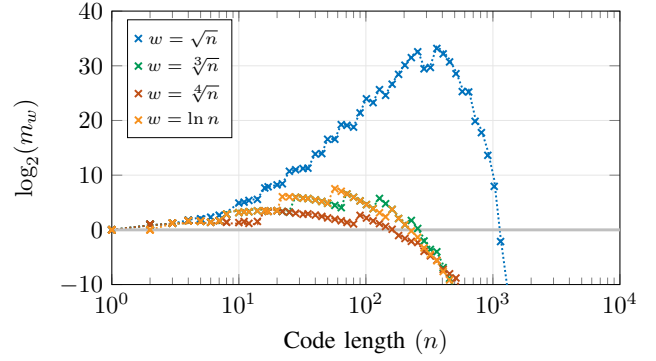
\begin{figure*}[t] 
    \centering
  \subfloat[{$v = \sqrt[4]{n}$}\label{fig:sqrt}]{%
           \centering
\resizebox{\columnwidth}{!}{
\begin{tikzpicture}
    \begin{axis}[
        xlabel={Code length $(n)$},
        ylabel={$\log_{2}(m_w)$},
        xmin=1, xmax=10e07,
        ymin=-500, ymax=3000,
        ytick={-1000, 0, 1000, 2000, 3000},
        legend pos = north west,
        legend style={font=\scriptsize},
        xmajorgrids=true,
        ymajorgrids=true,
        height  = 5.5cm,
        major grid style = {gray!20!white},
        width = \columnwidth,
        legend columns=1,
        xmode = log
    ]

\pgfplotstableread{Data/w_n_0.5.txt}{\vnine}
\addplot [RoyalBlue, thick, only marks, mark = x] table[x=X, y=Y] {\vnine};
\addlegendentry{$w = \sqrt{n}$};

\pgfplotstableread{Data/w_n_0.3.txt}{\vnine}
\addplot [ForestGreen, thick, only marks, mark = x] table[x=X, y=Y] {\vnine};
\addlegendentry{$w = \sqrt[3]{n}$};

\pgfplotstableread{Data/w_n_0.25.txt}{\vnine}
\addplot [Bittersweet, thick, only marks, mark = x] table[x=X, y=Y] {\vnine};
\addlegendentry{$w = \sqrt[4]{n}$};

\pgfplotstableread{Data/w_n_log_n.txt}{\vnine}
\addplot [BurntOrange, thick, only marks, mark = x] table[x=X, y=Y] {\vnine};
\addlegendentry{$w = \ln(n)$};

\pgfplotstableread{Data/w_n_0.5_smooth.txt}{\vnine}
\addplot [RoyalBlue, thick, densely dotted] table[x=X, y=Y] {\vnine};

\pgfplotstableread{Data/w_n_0.25_smooth.txt}{\vnine}
\addplot [Bittersweet, thick, densely dotted] table[x=X, y=Y] {\vnine};

\pgfplotstableread{Data/w_n_0.3_smooth.txt}{\vnine}
\addplot [ForestGreen, thick, densely dotted] table[x=X, y=Y] {\vnine};

\pgfplotstableread{Data/w_n_log_n.txt}{\vnine}
\addplot [BurntOrange, thick, densely dotted] table[x=X, y=Y] {\vnine};

\addplot [gray!50!white, very thick] coordinates{
(1,0)
(1e08, 0)
};

\end{axis}

\end{tikzpicture}
}}
    \hfill
  \subfloat[$v = \ln^2(n)$\label{fig:log}]{%
\resizebox{\columnwidth}{!}{
  \begin{tikzpicture}
    \begin{axis}[
        xlabel={Code length $(n)$},
        ylabel={$\log_{2}(m_w)$},
        xmin=1, xmax=1e04,
        ymin=-10, ymax=40,
        ytick={-10, 0, 10, 20, 30, 40},
        legend pos = north west,
        legend style={font=\scriptsize},
        xmajorgrids=true,
        ymajorgrids=true,
        height  = 5.5cm,
        major grid style = {gray!20!white},
        width = \columnwidth,
        legend columns=1,
        xmode = log
    ]

\pgfplotstableread{Data/v_log_sq_w_n_0.5.txt}{\vnine}
\addplot [RoyalBlue, thick, only marks, mark = x] table[x=X, y=Y] {\vnine};
\addlegendentry{$w = \sqrt{n}$};

\pgfplotstableread{Data/v_log_sq_w_n_0.3.txt}{\vnine}
\addplot [ForestGreen, thick, only marks, mark = x] table[x=X, y=Y] {\vnine};
\addlegendentry{$w = \sqrt[3]{n}$};

\pgfplotstableread{Data/v_log_sq_w_n_0.25.txt}{\vnine}
\addplot [Bittersweet, thick, only marks, mark = x] table[x=X, y=Y] {\vnine};
\addlegendentry{$w = \sqrt[4]{n}$};

\pgfplotstableread{Data/v_log_sq_w_n_log_n.txt}{\vnine}
\addplot [BurntOrange, thick, only marks, mark = x] table[x=X, y=Y] {\vnine};
\addlegendentry{$w = \ln{n}$};


\pgfplotstableread{Data/v_log_sq_w_n_0.5_smooth.txt}{\vnine}
\addplot [RoyalBlue, thick, densely dotted] table[x=X, y=Y] {\vnine};

\pgfplotstableread{Data/v_log_sq_w_n_0.3_smooth.txt}{\vnine}
\addplot [ForestGreen, thick, densely dotted] table[x=X, y=Y] {\vnine};

\pgfplotstableread{Data/v_log_sq_w_n_0.25_smooth.txt}{\vnine}
\addplot [Bittersweet, thick, densely dotted] table[x=X, y=Y] {\vnine};

\pgfplotstableread{Data/v_log_sq_w_n_log_n_smooth.txt}{\vnine}
\addplot [BurntOrange, thick, densely dotted] table[x=X, y=Y] {\vnine};

\addplot [gray!50!white, very thick] coordinates{
(1,0)
(1e04, 0)
};

\end{axis}

\end{tikzpicture}
}}

  \caption{Values of $\log_2(m_w)$ for $R = 0.8$ and several choices for $w$. 
In Figure \ref{fig:sqrt} we have $v = \sqrt[4]{n}$ while, in Figure \ref{fig:log}, we have $v = \ln^2(n)$.
In both figures, marks indicate the (logarithm of) EWD coefficients computed with the actual distribution, while dotted lines are used for the Bernoulli approximation.
Notice that lines get, in some cases, stepped because we had to round $n$, $r$ and $v$ to integers. This is more evident in Figure \ref{fig:log} because the figure reports smaller values of $n$ and $\log_2(m_w)$.
}
\label{fig:three graphs}
\end{figure*}

\begin{IEEEproof}
We first rewrite $\varepsilon(n, v, w)$ as 
$\Bigg(\left(1-\frac{2}{n/v}\right)^{n/v}\Bigg)^{\frac{vw}{n}}$.
Since $v = o(n)$, we have that $n/v$ goes to infinity as $n$ grows.
Then, we can use the following asymptotic expansion: for any $\alpha\in \mathbb R$ and growing $x$, we have
$\left(1-\frac{\alpha}{x}\right)^x\sim e^{-\alpha}$.
We then obtain
\begin{align*}
\varepsilon(n, v, w) &\nonumber \sim \left(e^{-2}\right)^{\frac{vw}{n}} = e^{\frac{-2vw}{n}}.
\end{align*}
Now: if $vw$ grows more than linearly with $n$, then we have that $e^{\frac{-2vw}{n}}$ goes to 0 for growing $n$ and, again, $\varepsilon(n, v, w) = o(1)$.

When $vw = \Theta(n)$, say, $vw = \alpha\cdot n$ for some $\alpha\in \mathbb R$, we then have $e^{\frac{-2vw}{n}}\sim \beta \in\mathbb R$.
However, we cannot have an EMD which is sublinear in $n$ since $\log_2(m_w)$ asymptotically grows as 
$$w\cdot \log_2(n/w)-n(1-R) \cdot \big(1-\log_2(1+\beta)\big).$$
Again, the second term grows faster than the first one hence in this regime we have $\log_2(m_w)<0$.

Let us now consider the case $vw = o(n)$.
In this case, we have $e^{\frac{-2vw}{n}} = o(1)$: since $e^{-x}\sim 1-x$ for $x \rightarrow 0$, we have $e^{\frac{-2vw}{n}}\sim 1-\frac{2vw}{n}$.
Then,
\begin{align*}
\log_2\left(1+\varepsilon(n,v,w)\right) &\nonumber \sim \log_2\left(1+1-\frac{2vw}{n}\right)\\\nonumber
& = 1+\log_2\left(1-\frac{vw}{n}\right)\\\nonumber
& \sim 1-\frac{vw}{n\cdot \ln(2)},
\end{align*}
where the last expansion is based on the fact that $\log_2(1-x)\sim \frac{-x}{\ln(2)}$ for $x \rightarrow 0$.
We then have
$$1-\log_2\big(1+\varepsilon(n, v, w)\big)\sim \frac{vw}{n\cdot \ln(2)}.$$
Considering this expansion into \eqref{eq:starting_point}, we get that $\log_2(m_w)$ grows as
\begin{align*}
w\cdot \log_2(n/w)-(1-R)\frac{vw}{\ln(2)}.
\end{align*}
Requiring $\log_2(m_w)\geq 0$ leads to 
$w\leq n\cdot 2^{\frac{-(1-R)v}{\ln(2)}}$.
\end{IEEEproof}
As a simple corollary, we get that, asymptotically, $m_w\geq 1$ with $w = o(n)$ can happen only if $v = O\big(\log(n)\big)$.
\begin{theorem}\label{the:log}
Let $r = (1-R)n$ for constant $R\in [0\hsp ; \hsp 1]$.
Then, existence of coefficients $m_w\geq 1$ for $w = o(n)$ can happen only if $v < \frac{\ln(2)}{1-R}\cdot \log_2(n)$.
\end{theorem}
\begin{IEEEproof}
We observe that, whenever $n\cdot 2^{\frac{-(1-R)v}{\ln(2)}}<1$, then there is no useful value $w$ which satisfies the requirements of Theorem \ref{the:final_theorem}.
Solving this inequality results in the thesis.
\end{IEEEproof}

\subsection{Some examples of EWD}

We would like to make some examples of the EWD arising from $\mathcal H_{n,r,v}$, in order to grasp the meaning of the theorems in the previous section.

In particular, we want to show that looking only at asymptotics to draw conclusions for how the EWD behaves, for concrete (but large) values of $n$, may sometimes be a bit misleading.
For instance, Theorem \ref{the:log} seems to imply that the only hope to have a code with codewords of low weight is that of using $v = O\big(\log(n)\big)$.
This is indeed true asymptotically but, for some choices of the parameters, the asymptotic regime may kick-in at very large values of $n$, say, much larger than those we require in many practical situations.

To show a practical example of this fact, let us take the case of $v = \sqrt[4]{n}$ and $R = 0.8$. 
Notice that, with this choice of $v$, we are not compliant with the thesis of Theorem \ref{the:log}, hence asymptotically all EWD coefficients for sublinear weights are $<1$.
However, for several (finite) values of $n$, we actually find that $m_w>1$ holds for many sublinear $w$.

To see some examples, consider Figure \ref{fig:sqrt}. 
As we can see, for all the considered choices of $w$, the value of $\log_2(m_w)$ becomes negative for sufficiently large values of $n$.
However, $\log_2(m_w)$ is always positive for a very large range of values (e.g., up to $10^6$).
This means that, in this case, the asymptotic regime kicks in rather late and, in practice, existence of codewords with very low weight happens even if the density does not satisfy the requirements of Theorem \ref{the:log}.

The same phenomenon can be observed in Figure \ref{fig:log} where we have used the same code rate but $v = \ln(n)^2$.
We can observe that, with respect to Figure \ref{fig:sqrt}, the values of $\log_2(m_w)$ are much smaller.
This is due to the fact, for the considered range of values for $n$, the distribution considered in Figure \ref{fig:log} outputs parity-check matrices which are denser than those considered in Figure \ref{fig:sqrt}.
Indeed, we have that $\ln(n)^2>\sqrt[4]{n}$ holds until $n\approx 10^{11}$.\footnote{To be precise, $\ln(n)^2$ and $\sqrt[4]{n}$ intersect for three values of $n$, which are approximately $0.41$, $3.18$ and $10^{11}$.}

\begin{figure}[H]
\resizebox{\columnwidth}{!}{
\begin{tikzpicture}
    \begin{axis}[
        xlabel={Code length $(n)$},
        ylabel={$\log_2(m_w)$},
        xmin=1, xmax=10e07,
        ymin=-10, ymax=200,
        ytick= {0, 50, 100, 150, 200},
        legend pos = north west,
        legend style={font=\scriptsize},
        xmajorgrids=true,
        ymajorgrids=true,
        height  = 5.5cm,
        major grid style = {gray!20!white},
        width = \columnwidth,
        legend columns=1,
        xmode = log
    ]

\pgfplotstableread{Data/comparison_v_n_05.txt}{\vnine}
\addplot [RoyalBlue, thick, only marks, mark = x] table[x=X, y=Y] {\vnine};
\addlegendentry{$v = \sqrt{n}$};

\pgfplotstableread{Data/comparison_v_n_1_3.txt}{\vnine}
\addplot [ForestGreen, thick, only marks, mark = x] table[x=X, y=Y] {\vnine};
\addlegendentry{$v = \sqrt[3]{n}$};

\pgfplotstableread{Data/comparison_v_n_025.txt}{\vnine}
\addplot [Bittersweet, thick, only marks, mark = x] table[x=X, y=Y] {\vnine};
\addlegendentry{$v = \sqrt[4]{n}$};

\pgfplotstableread{Data/comparison_v_n_log_n.txt}{\vnine}
\addplot [BurntOrange, thick, only marks, mark = x] table[x=X, y=Y] {\vnine};
\addlegendentry{$v = \ln(n)$};

\pgfplotstableread{Data/comparison_v_n_05_smooth.txt}{\vnine}
\addplot [RoyalBlue, thick, densely dotted] table[x=X, y=Y] {\vnine};

\pgfplotstableread{Data/comparison_v_n_1_3_smooth.txt}{\vnine}
\addplot [ForestGreen, thick, densely dotted] table[x=X, y=Y] {\vnine};

\pgfplotstableread{Data/comparison_v_n_025_smooth.txt}{\vnine}
\addplot [Bittersweet, thick, densely dotted] table[x=X, y=Y] {\vnine};

\pgfplotstableread{Data/comparison_v_n_log_n_smooth.txt}{\vnine}
\addplot [BurntOrange, thick, densely dotted] table[x=X, y=Y] {\vnine};

\addplot [gray!50!white, very thick] coordinates{
(1,0)
(1e08, 0)
};

\end{axis}

\end{tikzpicture}
}
\caption{Values of $\log_2(m_w)$ for $R = 0.8$, $w = \ln(n)$ and several values of $v$.}
\label{fig:several_v}
\end{figure}
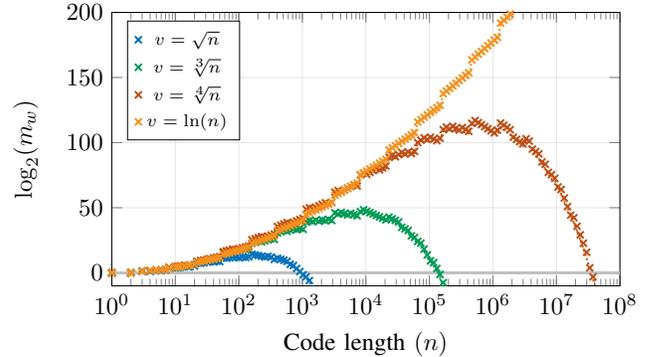

Finally, in Figure \ref{fig:several_v} we consider how the values of $\log_2(m_w)$ behave as $v$ changes; for this comparison, we have used $w = \ln(n)$.
As we expected, the values get higher as parity-check matrices get sparser, i.e., as $v$ decreases.
Observe that for $v = \ln(n)$ we have $v<\frac{\ln(2)}{(1-R)}\log_2(n)$, as requested by Theorem \ref{the:log} and, indeed, the values of $\log_2(m_w)$ are always positive, regardless of $n$.

\section{The Algorithm for Sampling Self-orthogonal Sparse Matrices}
\label{sec:isd}

In this section we describe and analyze the algorithm to sample self-orthogonal parity-check matrices.

\subsection{The algorithm}
The algorithm is pretty similar to that from \cite{albrecht2025hollow} but there are some differences.
First, in this section we only consider the binary case and this simplifies a bit the analysis.
\footnote{For instance, we only deal with linear systems. In any case, with some effort, our algorithm can be generalized to non binary fields, if needed. Later on, we will comment about how this can be done.}
Moreover, since we are not interested in formally proving that the sampled parity-check matrix is uniformly random over the co-domain, we avoid the rejection sampling used in \cite{albrecht2025hollow}.
As a main difference, though, here we need to sample low weight codewords.
We are going to do this using ISD.

The procedure is described in Algorithm \ref{alg:sample}.
For each $u\in\{1,\hsp\cdots,\hsp r\}$, we assume we already know $u-1$ linearly independent vectors $\{\6h_1,\hsp\cdots,\hsp\6h_{u-1}\}$, all with weight $v$ and such that $\6h_j\6h_t^\top = 0$, for each pair of indices $j$ and $t$. 
Now, we want to find another vector $\6h_u\in\mathbb F_2^n$, having weight $v$, which must be orthogonal to each of the already found vectors $\6h_j$.
We can do this by searching for weight-$v$ codeword in the linear code whose parity-check matrix is defined by equations $\6h_1,\hsp \cdots,\hsp \6h_{u-1}$.
Actually, since we want also self-orthogonality for $\6h_u$, we can add as additional parity-check equation the all ones vector: this enforces that the codeword must have even Hamming weight.

\begin{algorithm}[ht]
\KwIn{$r,n,v\in \mathbb N$ with $r,v\leq n$ and $v$ even}
\KwOut{$\6H\in\mathbb F_2^{r\times n}$ such that $\6H\6H^\top = \60$ and all rows have weight $v$}
\vspace{0.5em}

\tcp{Sample first codeword}
Sample $\6h_1\in\mathbb F_2^n$ with weight $v$\;
\vspace{2mm}

Set $u = 1$;\tcp{Counter for number of found vectors}
\While{$u < r$}{
\vspace{2mm}

\tcc{Set parity-check matrix for $\code_u$}
Set $\6H_u = \begin{pmatrix}\6h_1\\\vdots \\ \6h_{u}\\ (1,\hsp\cdots,\hsp 1)
\end{pmatrix}$\;
\vspace{2mm}

\tcp{Call ISD to find weight-$v$ codeword in $\code_u$}
Repeatedly call ISD, until a codeword $\6c \in\mathrm{Ker}(\6H_u)$ with weight $v$ is found \;
\vspace{2mm}

\tcp{If $\6c\not\in\code_u$, enrich the found basis and increase $u$; otherwise, continue searching for a codeword in $\code_u$}
\If{$\6c\not\in \mathrm{Span}(\6h_1,\hsp\cdots,\hsp \6h_u)$}{
Update $u \gets u+1$\;
Set $\6h_u = \6c$\;
}
}
\Return $\6H = \begin{pmatrix}\6h_1\\\vdots \\ \6h_r\end{pmatrix}$\;
\caption{$\mathsf{SampleDualContaining}$}
\label{alg:sample}
\end{algorithm}

By doing this, we obtain the code $\code_u\subseteq \mathbb F_2^n$ whose parity-check matrix is $\6H_u$ (see line 4 of the algorithm).
Observe that $\code_u$ has length $n$ and dimension which is either $n-(u-1)$, if the all ones vector is contained in the span of $\6h_1,\hsp\cdots,\hsp\6h_{u-1}$, and $n-u$ otherwise.
Then, we search for a low weight codeword $\6c\in \code_u$ and, whenever we find it, we check whether it is in the span of $\6h_1,\hsp \cdots,\hsp \6h_{u-1}$: if not, we set $\6h_u = \6c$ and proceed with step $u+1$, otherwise we start searching for a new codeword.

Notice that, for $u = 1$, we just need to sample from the Hamming sphere with the desired radius.

\subsection{Choice of $v$}

To make the algorithm work, we need to choose a weight $v$ such that, at every step, the code $\code_u$ contains codewords of weight $v$ with large probability.
As we shall see later on, the value of $v$ has also an impact on the time complexity of the algorithm.

To accomplish this, we need to model the weight distribution of the sequence of codes $\code_1,\hsp \code_2,\hsp \cdots$.
A formal derivation is going to be definitely non trivial, since this would require to a) take into account self-orthogonality (which necessarily imposes some dependence between the rows of the parity-check matrix), and b) consider that the codes $\code_1,\hsp \code_2,\hsp\cdots$ are obviously correlated.
We will simplify the theoretical model, relying on the following heuristic.
\begin{heuristic}\label{heu:ewd}
For each $u\in\{1,\hsp \cdots,\hsp r\}$, we treat $\begin{pmatrix}\6h_1\\\vdots\\\6h_{u-1}\end{pmatrix}$ as a sample from $\mathcal H_{n,u-1,v}$.
\end{heuristic}
This allows us to study the weight distribution of each code $\code_u$ through the analysis from Section \ref{sec:distributions}.
Indeed, we observe that $\code_u$ is the code obtained by taking only even weight codewords from the code whose parity-check matrix is formed by the equations $\6h_1,\hsp\cdots,\hsp\6h_{u-1}$.
Since by design we have that $v$ is even, we derive the expected number of weight-$v$ codewords in $\code_u$ by relying on Heuristic \ref{heu:ewd} and consequently using Theorem \ref{the:ewd_distrib}, taking $n$ as code length and $u-1$ as code redundancy.

Let us indicate this number as $m_v^{(u)}$; then
$$m_v^{(u)} = \binom{n}{v}\rho_{v}^{u-1}.$$
To guarantee that each call to ISD can find codewords with the desired weight with large probability, we should have $m_v^{(u)}\geq 1$ for each value of $u$.

Actually, observe that $m_v^{(u)}$ decreases as $u$ increases.
Hence, we actually just need to guarantee this condition for the largest value of $u$, that is, $u = r$.
In other words, we recommend choosing $v$ such that 
\begin{equation}
\label{eq:choice_of_v}
m_v^{(r)} = \binom{n}{v}\rho_v^{r-1}\geq 1.
\end{equation}
Actually, we recommend to choose $v$ so that $m_v^{(r)}\gg 1$.
Indeed, as a rule of thumb, the larger this coefficient, the higher the probability that the code indeed contains codewords of weight $v$.

Asymptotically, according to Theorem \ref{the:log}, we shall use $v<\frac{\ln(2)}{1-R}\cdot \log_2(n)$.

\subsection{Tweaks to ISD}

We consider some tweaks to boost the performances of ISD. The first one is due to the sparsity of the parity-check matrix, which interferes with the distribution of information sets; the second one, instead, is due to the fact that for the early values of $u$, we are searching for sparse codewords in a code with very small redundancy.
\medskip

We consider, as input to ISD, the matrix $\6H_u$ with dimension $u\times n$. 
Observe that, if $(1,\hsp\cdots,\hsp 1)\in\mathrm{Span}(\6h_1,\hsp\cdots,\6h_{u-1})$, we should actually consider that it has $u-1$ rows: using $u$ or $u-1$ makes little difference here.
\medskip

\paragraph*{\textbf{Choosing the information set}}
Since the matrix $\6H_u$ is very sparse, the probability that a random set of $u$ columns yields a non singular matrix may be smaller than $\gamma_2$, i.e., the probability that a random $r\times r$ binary matrix is non singular (see e.g. \cite{blomer1997rank}).
To avoid that the first part of ISD (permutation and Gaussian elimination) is repeated too many times, we modify the algorithm so that, after $\6P\in S_n$ is sampled uniformly at random, we search for an information set in $\6H' = \6H\cdot \6P$.
This can be done efficiently by, e.g., using Gaussian elimination to put $\6H'$ in Reduced Row Echelon Form (RREF) and then using the coordinates of non pivoted columns as the information set.
Let $\6P_{\mathrm{RREF}} \in S_n$ be a permutation that moves these coordinates to the rightmost positions.
Then, we set $\6H'' = \6H'\cdot \6P_{\mathrm{RREF}}= (\6H''_1,\hsp \6H''_2)$, with $\6H''_2$ being the $r\times r$ matrix formed by the rightmost $r\times r$ columns: by design, this is non singular.
We then put $\6H''$ in systematic form $(\6A,\hsp \6I_r)$ with $\6A = \6H''^{-1}_{2}\cdot \6H''_1$, and we are ready for the enumeration phase.
\medskip

\paragraph*{\textbf{Choosing $p$}}

When the redundancy is very small, there is a very simple way to select the value of $p$.
Indeed, after enumeration on the information set, we have to check the weight on the non information set part.
On average, this will be given by $u/2$.
Thus, if $u/2 < v$, we can just choose $p\approx v-u/2$; obviously, if $u$ is not even, we will need to round this value.
This guarantees that, with large probability, we find a codeword with the desired weight.
Indeed, treating $\6c''$ as a random vector of length $u$, the probability to get the desired weight is
$\binom{u}{u/2}2^{-u}$.
Using the well known approximation for the central binomial coefficient, we get that this probability is approximately $1/\sqrt{\pi u/2}$.
Thus, we find a codeword after approximately $\sqrt{\pi u/2}$ attempts.

When instead $u$ gets larger (say, $u>2v$), we choose $p$ aiming to optimize the time complexity.
In practice, we will set $p$ as a small constant (say, $p = 2$ or $p = 3$) in order to amortize the computational costs in each iteration and making the enumeration phase, approximately, as costly as Gaussian elimination.

\subsection{Complexity}
\label{subsec:isd_polynomial}

We first estimate the time complexity of Algorithm \ref{alg:sample}.
\begin{proposition}\label{prop:cost_sample}
Let $r,n,v\in\mathbb N$ such that $r< n/2$, and $v$ even satisfies Eq. \eqref{eq:choice_of_v}. Then, under heuristic \ref{heu:codewords_w}, using $\leb$ as ISD, the time complexity of Algorithm \ref{alg:sample} can be estimated as
\begin{equation*}
\sum_{u = 1}^{r-1}\frac{T_{\leb}(n,\hsp n-u-1)}{(1-2^{2(u+1)-n})\cdot P_{\leb}\left(n, \hsp n-u-1, \hsp v, \hsp m_v^{(u)}\right)}.
\end{equation*}
\end{proposition}
\begin{IEEEproof}
For simplicity, we assume that in the $u$-th step the code $\code_u$ has always redundancy $u+1$, i.e., that $(1,\cdots,1)\not\in\mathrm{Span}(\6h_1,\cdots,\6h_{u-1})$.
For each step, the cost is dominated by the cost of repeatedly call ISD.
Summing over all steps and using $m_v^{(u)}$ as an estimate for the amount of codewords in $\code_u$, recalling Proposition \ref{prop:lee}, we get that in the $u$-th step the cost to sample a weight-$v$ codeword in $\code_u$ can be estimated as $T_{\leb}(n,\hsp n-u-1)/ P_{\leb}(n, \hsp n-u-1, \hsp v, \hsp m_v^{(u)})$.

We now consider that, in the $u$-th step, we need to sample a codeword which is not linearly dependent with $\6h_1,\hsp\cdots,\hsp\6h_{u-1}$.
Indeed, by design we have $\code_u^\bot \subseteq \code_u$ so there is a chance the codeword sampled by ISD is actually in $\code_u^\bot$.
In other words, a weight-$v$ codeword is valid only if it is coming from $\code_u\setminus \code_u^\bot$.
Under the simplifying assumption that weight-$v$ codewords are uniformly distributed in the code, we can estimate such a probability as
\begin{align*}
\frac{|\code_u\setminus \code_u^\bot|}{|\code_u|} &\nonumber = \frac{2^{n-u-1}-2^{u+1}}{2^{n-u-1}} = 1-2^{2(u+1)-n}.
\end{align*}
The reciprocal of the above probability gives the average amount of weight-$v$ codewords we need to find, for each value of $u$, before a codeword enriching the basis is found.
\end{IEEEproof}
\begin{remark}
The above proposition is given only for $ r < n/2$ because, for $r = n/2$ (or, more generally, when $r$ is pretty close to $n/2$), the analysis would become rather sketchy.
Indeed, whenever $u = n/2-1$, we have that $\code_u$ has either dimension $n/2$ or $n/2+1$, depending on whether $(1,\hsp\cdots,\hsp 1)\in\mathrm{Span}(\6h_1,\hsp \cdots, \hsp \6h_{u-1})$ or not.
Notice that, if it has dimension $n/2$, then our algorithm is expected to halt with very low probability.
Indeed, in this case we have $\code_u = \code_u^\bot$ hence each codeword we sample is of the form $a(1,\hsp \cdots,\hsp 1)+ \sum_{i = 1}^{u-1}b_i \6h_i$, where  $a, b_1,\cdots,b_{u-1}\in\mathbb F_2$.
Whenever $a = 0$, then we get exactly a linear combination of $\6h_1,\hsp\cdots,\hsp\6h_{u-1}$.
The only possibility is to have $a = 1$ and $\mathrm{wt}\left(\sum_{i = 1}^{u-1}b_i \6h_i\right) = n-v$.

\end{remark}
Looking closely into the terms that contribute to the complexity of Proposition \ref{prop:cost_sample}, we observe that when $p$ is a small constant, then the term $T_{\leb}(n,\hsp n-u)$ is a polynomial in $n$.
Moreover, for any $u\leq n/2 - 2$ we have
$\frac{1}{1-2^{2(u+1)-n}}\leq \frac{1}{1-2^{-2}} = 4/3$.

The only term that can become exponential is $1/P_{\leb}\left(n,\hsp n-u,\hsp v,\hsp m_v^{(u)}\right)$, which estimates the average number of ISD calls we need to find a codeword with weight $v$.
However, assuming heuristics \ref{heu:info_set} and \ref{heu:codewords_w} hold also in our interest case (i.e., when studying ISD for sparse parity-check matrices), then we can show that in many cases this tern is a polynomial of $n$, as well.

First, recalling \cite{canto2016analysis}, the cost of any binary ISD for a sublinear weight $v$ and a family of codes with constant code rate $R$, when there is a unique codewod with the desired weight, asymptotically is 
$$2^{-v\cdot \log_2(1-R)\cdot \big(1+o(1)\big)}.$$
Since asymptotically we have $v = O\big(\log_2(n)\big)$, the above formula yields a polynomial of $n$.

Moreover, we should also consider that when the amount of weight-$v$ codewords is very large, then the average number of ISD calls before a valid codeword is found is reduced drastically.
Indeed, let us focus on $\leb$: recalling Proposition \ref{prop:lee} and using $m_v$ as in Theorem \ref{the:ewd_distrib} to estimate the average number of codewords with weight $v$ we have in the last step of the algorithm, the success probability of one call can be estimated as\footnote{For the sake of simplicity, we consider a code with redundancy $r = n-k$ and, consequently, raise $\rho_v$ to the power of $n-k$.
According to the analysis in the previous section, it should be instead $\rho_v^{n-k-1}$. Since we are looking at asymptotics, having a $\pm 1$ difference does not make a difference.}
$$\min\left\{1\hsp; \hsp\binom{k}{p}\binom{n-k}{v-p}\cdot \rho_v^{n-k}\right\}.$$
Whenever $\binom{k}{p}\binom{n-k}{v-p}\cdot \rho_v^{n-k}\geq 1$, then the average number of iterations before a valid codeword is returned is estimated as 1.
In many situations, this is indeed the case.

For instance, let us consider $\leb$ with parameter $p$ chosen as a small constant (say, $p = 2$ or $p = 3$).
Whenever $v$ is constant, asymptotics show that indeed $\binom{k}{p}\binom{n-k}{v-p}\cdot \rho_v^{r}\geq 1$.
Since for constant $p$ we have $\binom{k}{p} = O (n^p)$, the worst case cost of each iteration is $O\left(\max\left\{n^3\hsp;\hsp n^{p+1}\right\}\right)$\footnote{This is a worst case because we do not consider that, on average, a valid codeword is found before all weight-$p$ vectors are enumerated.}.
We need to call ISD for $r$ times so, overall, we get a worst case cost for our algorithm which is
$$O\left(\max\left\{n^4\hsp;\hsp n^{p+2}\right\}\right).$$
As we show in the next sections, in our simulations we actually observed that, for the majority of experiments, one ISD call is enough to find the desired codeword.
\medskip

All in all, there are strong evidences that our sampling algorithm runs in polynomial time and making such a claim is indeed rather tempting.
The above analysis, however, shall be taken with a grain of salt.
Indeed, the assessment of ISD is based on some heuristics which, in our case, may not hold.

In particular, the parity-check matrices we are dealing with are rather sparse and this may affect the distribution of information sets.
Thus, we cannot guarantee that information sets are equally distributed over codewords.
In other words, for some codewords the success probability may be lower than what claimed in Proposition \ref{prop:lee} because there are very few information sets that overlap with the support of the codeword in exactly $p$ positions.
While this phenomenon is true also for random codes, we believe in our case it may get amplified because of sparsity.
Thus, Heuristic \ref{heu:codewords_w} may not hold in our case and, consequently, the success probability shall be estimated in a way which is different than Proposition \ref{prop:lee}.

Our numerical experiments confirm that the above phenomenon indeed happens, even though (as one expects) the impact is extremely mild.
However, we feel this is already enough to refrain us from making claims that may seem too bold without a solid, theoretical back-up.
We thus leave such an analysis for future works.
We also highlight that, instead of just analyzing how standard ISD would behave for sparse parity-check matrices, it would probably make more sense to design an ISD algorithm that is specifically tailored for sparse matrices.

\subsection{Getting rid of unlucky columns}
\label{subsec:unlucky}

By design, we are not able to guarantee that a matrix sampled by Algorithm \ref{alg:sample} satisfies certain properties which would make it a parity-check matrix for a ``good'' LDPC code.
However, we show how one can modify the matrix by eliminating ``unlucky'' columns. 

As an example, we consider columns with very few set entries.
Modeling $\6H\in\mathbb F_2^{r\times n}$ as a sample of $\mathcal H_{n,r,v}$, the expected amount of of columns with weight $z$ is
\begin{equation}
\label{eq:binomial_columns}
t_z = n\cdot \binom{r}{z}\left(\frac{v}{n}\right)^z\left(1-\frac{v}{n}\right)^{r-z}.
\end{equation}
For instance, the expected amount of null columns is
$$t_0 = n\cdot \left(1-\frac{v}{n}\right)^r.$$
For $r = (1-R)n$, we have
\begin{align*}
t_0 = n\cdot \Bigg(\underbrace{\left(1-\frac{v}{n}\right)^\frac{n}{v}}_{\sim e^{-1}}\Bigg)^{v(1-R)} \sim n\cdot e^{-v(1-R)}.
\end{align*}
If $v$ is constant, then this is linear in $n$ while, when $v = \alpha\log_2(n)$ for some $\alpha>0$, this grows as $n^{1-\alpha(1-R)}$.

Such columns can be eliminated from $\6H$ (technically, we puncture $\code^\bot$ in the corresponding coordinates), and get a new matrix $\6H'$ which is a parity-check matrix for a new code $\code'$ with length $n'$ and same redundancy $r$.
Notice that self-orthogonality is trivially guaranteed and, moreover, the new length $n'$ is not much smaller than the initial one.
Indeed, the expected value for $n'$ is $n-t_0$ and, for $v = o(n)$, this is
$$n' \sim n\cdot\left(1-e^{-v(1-R)}\right).$$
For constant $v$, $n'$ is linear in $n$ while, whenever $v$ grows with $n$, the value of $n'$ converges to $n$ since $e^{-v(1-R)}$ goes to 0.

We now describe how to get rid of columns with weight $\geq 1$.
Let $J\subseteq \{1,\hsp \cdots,\hsp n\}$ be the set of columns that shall be removed, and $E\subseteq \{1,\hsp \cdots,\hsp r\}$ the set of rows so that, for every coordinate $j\in J$ and every $e\in E$, we have $h_{e,j} = 0$.
Then, we choose $\6H'\in\mathbb F_2^{|E|\times (n-|J|)}$ as the submatrix formed by rows indexed by $E$ and columns which are not indexed by $J$. 
Self-orthogonality is trivially preserved.

For instance, let us assume we want to get rid of all columns with weight $1$.
Notice that, in this case, we have $|E|\leq |J|$.
Asymptotically, for $v = o(n)$, we have
\begin{align*}
t_1 & \nonumber = n\cdot r\cdot \left(\frac{v}{n}\right)\left(1-\frac{v}{n}\right)^{r-1}\\\nonumber
& = n\cdot v \cdot (1-R)\underbrace{\left(1-\frac{v}{n}\right)^{r}}_{\sim e^{-v(1-R)}} / \underbrace{\left(1-\frac{v}{n}\right)}_{\sim 1}\\\nonumber
& \sim n\cdot v \cdot (1-R)\cdot e^{-v(1-R)}.
\end{align*}
Again, this is very small with respect to $n$. For constant $v$, this is linear in $n$ while, if $v = \log_2(n)$, we have $t_1\sim v\cdot n^{-R}\cdot \log_2(n)$.

\section{Numerical validation and results}
\label{sec:results}
In this section we present some numerical validations and results.
All the code we have used for the simulations can be downloaded from 
\begin{center}
\url{https://github.com/psantini1992/DualContaining_LDPC}.
\end{center}

\subsection{Validation of EWD}

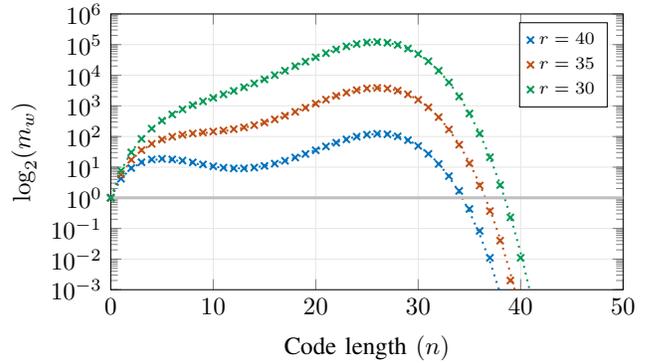
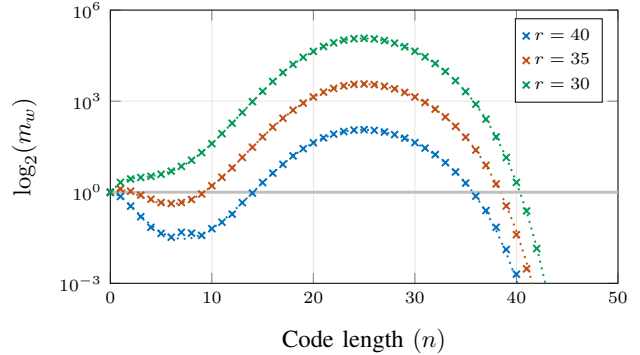
\begin{figure}[h] 
    \centering
  \subfloat[{$v = 3$}\label{fig:n50_v3}]{%
           \centering
\resizebox{\columnwidth}{!}{
\begin{tikzpicture}
    \begin{axis}[
        xlabel={Code length $(n)$},
        ylabel={$\log_2(m_w)$},
        xmin=0, xmax=50,
        ymin=1e-03, ymax=1000000,
        ytick={0.001, 0.01, 0.1, 1, 10, 100, 1000, 1e04, 1e05, 1e06},
        legend pos = north east,
        legend style={font=\scriptsize},
        xmajorgrids=true,
        ymajorgrids=true,
        major grid style = {gray!20!white},
        height = 5.5cm,
        width = \columnwidth,
        legend columns=1,
        ymode = log
    ]

\pgfplotstableread{Data/emp_wef_n_50_r_40_v_3.txt}{\vnine}
\addplot [RoyalBlue, thick, only marks, mark = x] table[x=X, y=Y] {\vnine};
\addlegendentry{$r = 40$};

\pgfplotstableread{Data/emp_wef_n_50_r_35_v_3.txt}{\vnine}
\addplot [Bittersweet, thick, only marks, mark = x] table[x=X, y=Y] {\vnine};
\addlegendentry{$r = 35$};

\pgfplotstableread{Data/emp_wef_n_50_r_30_v_3.txt}{\vnine}
\addplot [ForestGreen, thick, only marks, mark = x] table[x=X, y=Y] {\vnine};
\addlegendentry{$r = 30$};

\pgfplotstableread{Data/th_wef_n_50_r_40_v_3.txt}{\vnine}
\addplot [RoyalBlue, thick, dotted] table[x=X, y=Y] {\vnine};

\pgfplotstableread{Data/th_wef_n_50_r_35_v_3.txt}{\vnine}
\addplot [Bittersweet, thick, dotted] table[x=X, y=Y] {\vnine};

\pgfplotstableread{Data/th_wef_n_50_r_30_v_3.txt}{\vnine}
\addplot [ForestGreen, thick, dotted] table[x=X, y=Y] {\vnine};

\addplot [gray!50!white, very thick] coordinates{
(0, 1)
(50, 1)
};

\end{axis}

\end{tikzpicture}
}}
    \hfill
  \subfloat[$v = 5$\label{fig:n50_v5}]{%
\resizebox{\columnwidth}{!}{
  \pgfplotsset{every tick label/.append style={font=\scriptsize}}

\begin{tikzpicture}
    \begin{axis}[
        xlabel={Code length $(n)$},
        ylabel={$\log_{2}(m_w)$},
        xmin=0, xmax=50,
        ymin=1e-03, ymax=1000000,
        ytick={0.001, 1, 1000, 1e06},
        legend pos = north east,
        legend style={font=\scriptsize},
        xmajorgrids=true,
        ymajorgrids=true,
        yminorgrids =true,
        height  = 6cm,
        major grid style = {gray!20!white},
        legend columns=1,
        height = 5.5cm,
        width = \columnwidth,
        ymode = log
    ]

\pgfplotstableread{Data/emp_wef_n_50_r_40_v_5.txt}{\vnine}
\addplot [RoyalBlue, thick, only marks, mark = x] table[x=X, y=Y] {\vnine};
\addlegendentry{$r = 40$};

\pgfplotstableread{Data/emp_wef_n_50_r_35_v_5.txt}{\vnine}
\addplot [Bittersweet, thick, only marks, mark = x] table[x=X, y=Y] {\vnine};
\addlegendentry{$r = 35$};

\pgfplotstableread{Data/emp_wef_n_50_r_30_v_5.txt}{\vnine}
\addplot [ForestGreen, thick, only marks, mark = x] table[x=X, y=Y] {\vnine};
\addlegendentry{$r = 30$};

\pgfplotstableread{Data/th_wef_n_50_r_40_v_5.txt}{\vnine}
\addplot [RoyalBlue, thick, dotted] table[x=X, y=Y] {\vnine};

\pgfplotstableread{Data/th_wef_n_50_r_35_v_5.txt}{\vnine}
\addplot [Bittersweet, thick, dotted] table[x=X, y=Y] {\vnine};

\pgfplotstableread{Data/th_wef_n_50_r_30_v_5.txt}{\vnine}
\addplot [ForestGreen, thick, dotted] table[x=X, y=Y] {\vnine};

\addplot [gray!50!white, very thick] coordinates{
(0, 1)
(50, 1)
};

\end{axis}

\end{tikzpicture}
}}

  \caption{Comparison between theoretical and empirical EWD. The empirical EWD has been estimated by averaging the weight distributions of $10^4$ codes.
In both figures, markers are used for the empirical EWD, while dotted lines are employed for the theoretical EWD.
For both figures, we have considered $n = 50$; in Figure \ref{fig:n50_v3} we have $v = 3$, in Figure \ref{fig:n50_v5} instead $v = 5$.
}
\label{fig:validation}
\end{figure}

First, to validate the theoretical derivation of the EWD for the distribution $\mathcal H_{n,r,v}$, we have averaged the weight enumerators for several codes sampled from $\mathcal H_{n,r,v}$ and have compared with the theoretical expression for the EWD from Theorem \ref{the:ewd_distrib}. 
Since computing the weight distribution is rather time-consuming, we have considered only codes with dimension not greater than $20$.
The comparison is shown in Figure \ref{fig:validation}.
We see that empirical and theoretical values match pretty well.
We observe that this was just a sanity check since Theorem \ref{the:ewd_distrib} does not depend on any heuristic.

\subsection{Validation of ISD}

To validate the use of ISD to find low weight codewords, we have ran experiments on some codes; here, we present a selection of these experiments, for $n = 80$, $r = 40$ and $v = 7$.
For these parameters, the GV distance is $d = 10$.
We used ISD to find codewords with weight $4\leq w \leq 10$.
We wanted to verify the amount of distinct codewords we were able to find.
To do this, for each value of $w$, we have generated 100 codes and then, for each code, called ISD for 100 times. 
We have measured both the average number of ISD calls to find a codeword with weight $w$ is found, as well as the number of distinct codewords we have been able to produce.
We have repeated the experiment using two configurations for $\leb$, namely, $p = 2$ and $p = 3$.

In Table \ref{tab:validation_isd} we compare theoretical and empirical estimates.
We observe there is a slight difference between theoretical and empirical values and, in particular, theoretical values are always slightly larger than empirical ones.
This is coherent with what we discussed in Section \ref{subsec:isd_polynomial}.
In any case, the empirical results match pretty well with the theoretical estimates as the difference is rather mild.

As an explanation, we believe for some codewords there are very few information sets that intercept its support in exactly $p$ positions.
As we have already argued, this is due to the sparsity of the parity-check matrix.
In other words, ISD is going to have a very hard time finding such ``unlucky'' codewords and, with somewhat large probability, will not return one of them.
This is even supported by the fact that the average number of iterations to find a codeword is (slightly) larger than the theoretical prediction.
\begin{table*}[t]
\renewcommand{\arraystretch}{1.2}
\resizebox{\textwidth}{!}{
\begin{tabular}{lc|cc|cc|cc|cc|cc|cc|cc|}
\cline{3-16}&                      & \multicolumn{2}{c|}{$w = 4$}            & \multicolumn{2}{c|}{$w = 5$}            & \multicolumn{2}{c|}{$w = 6$}            & \multicolumn{2}{c|}{$w = 7$}            & \multicolumn{2}{c|}{$w = 8$}              & \multicolumn{2}{c|}{$w = 9$}              & \multicolumn{2}{c|}{$w = 10$}             \\ \cline{3-16} & \multirow{-2}{*}{}   & Th.    & Emp.   & Th.    & Emp.   & Th.    & Emp.   & Th.    & Emp.   & Th.     & Emp.    & Th.     & Emp.    & Th.     & Emp.    \\ \hline
\multicolumn{1}{|l|}{}       & Value of $m_w$                & \cellcolor[HTML]{EFEFEF}$3.69$ & $2.23$ & \cellcolor[HTML]{EFEFEF}$4.59$ & $2.19$ & \cellcolor[HTML]{EFEFEF}$6.35$ & $3.74$ & \cellcolor[HTML]{EFEFEF}$9.93$ & $5.87$ & \cellcolor[HTML]{EFEFEF}$17.70$ & $11.66$ & \cellcolor[HTML]{EFEFEF}$35.73$ & $24.89$ & \cellcolor[HTML]{EFEFEF}$80.86$ & $51.43$ \\ 
\multicolumn{1}{|l|}{\multirow{-2}{*}{\rotatebox[origin=c]{90}{$p = 2$}}} & Avg. number of ISD calls & \cellcolor[HTML]{EFEFEF}$1.20$ & $2.09$ & \cellcolor[HTML]{EFEFEF}$1.20$ & $2.77$ & \cellcolor[HTML]{EFEFEF}$1.22$ & $2.15$ & \cellcolor[HTML]{EFEFEF}$1.21$ & $2.14$ & \cellcolor[HTML]{EFEFEF}$1.17$  & $1.99$  & \cellcolor[HTML]{EFEFEF}$1.11$  & $1.56$  & \cellcolor[HTML]{EFEFEF}$1.05$  & $1.33$  \\ \hline
\multicolumn{1}{|l|}{}                                                    & Value of $m_w$                & \cellcolor[HTML]{EFEFEF}$3.69$ & $2.72$ & \cellcolor[HTML]{EFEFEF}$4.59$ & $3.71$ & \cellcolor[HTML]{EFEFEF}$6.35$ & $5.41$ & \cellcolor[HTML]{EFEFEF}$9.93$ & $8.67$ & \cellcolor[HTML]{EFEFEF}$17.70$ & $15.30$ & \cellcolor[HTML]{EFEFEF}$35.73$ & $33.97$ & \cellcolor[HTML]{EFEFEF}$80.86$ & $63.95$ \\ 
\multicolumn{1}{|l|}{\multirow{-2}{*}{\rotatebox[origin=c]{90}{$p = 3$}}} & Avg. number of ISD calls & \cellcolor[HTML]{EFEFEF}$1.53$ & $2.17$ & \cellcolor[HTML]{EFEFEF}$1.20$ & $1.83$ & \cellcolor[HTML]{EFEFEF}$1.09$ & $1.65$ & \cellcolor[HTML]{EFEFEF}$1.04$ & $1.37$ & \cellcolor[HTML]{EFEFEF}$1.01$  & $1.22$  & \cellcolor[HTML]{EFEFEF}$1.00$  & $1.08$  & \cellcolor[HTML]{EFEFEF}$1.00$  & $1.03$  \\ \hline
\end{tabular}
}
\caption{Comparison between theoretical results, for codes with $n = 80$, $r = 40$ and $v = 7$. We have generated 100 codes and, for each of them, called $\leb$ for 100 times.}
\label{tab:validation_isd}
\end{table*}

\begin{table*}[t]
\renewcommand{\arraystretch}{1.2}
\centering
\begin{tabular}{|ccc|c|c|cc|}
\hline
\multicolumn{3}{|c|}{Code parameters}                       & \multirow{2}{*}{\begin{tabular}[c]{@{}c@{}}GV \\ distance\end{tabular}} & \multirow{2}{*}{\begin{tabular}[c]{@{}c@{}}Value of \\ $m_v^{(r)}$\end{tabular}} & \multicolumn{2}{c|}{Benchmark results}                                     \\ \cline{1-3} \cline{6-7} 
\multicolumn{1}{|c|}{$n$} & \multicolumn{1}{c|}{$r$} & $v$  &                                                                         &                                                                            & \multicolumn{1}{c|}{Average time (seconds)} & Avg. number of ISD calls \\ \hline\hline
\multirow{5}{*}{250}      & \multirow{5}{*}{80}     & $6$  & \multirow{5}{*}{16}                                                     & $5.25\cdot 10^6$    & \multicolumn{1}{c|}{$2.13$} & $1$     \\
&  & $8$  &      & $2.69\cdot 10^6$   & \multicolumn{1}{c|}{$2.31$} & $1$\\
   &   & $10$ &    &     $3.82\cdot 10^5$ & \multicolumn{1}{c|}{$3.62$}  & $1$ \\ 
   &   & $12$ &    &     $4.52\cdot 10^4$ & \multicolumn{1}{c|}{$22.84$}  & $1$ \\ 
   &   & $14$ &    &     $1.22\cdot 10^4$ & \multicolumn{1}{c|}{$412.19$}  & $1.0025$ \\ 
   \hline
\multirow{3}{*}{500}      & \multirow{3}{*}{200}     & $6$  & \multirow{3}{*}{41}                                                     & $1.56\cdot 10^7$    & \multicolumn{1}{c|}{$22.77$}                & $1$     \\
&  & $8$  &      &   $2.01\cdot 10^6$ & \multicolumn{1}{c|}{$25.40$} & $1.0005$\\
   &   & $10$ &    &    $1.85\cdot 10^4$ & \multicolumn{1}{c|}{$164.05$} & 1.0010 \\ \hline
\multirow{3}{*}{1000}      & \multirow{3}{*}{400}     & $6$  & \multirow{3}{*}{81}                                                     & $8.77\cdot 10^8$    & \multicolumn{1}{c|}{$178.99$}    & $1$     \\
&  & $8$  &      &  $3.07\cdot 10^8$  & \multicolumn{1}{c|}{$235.01$} & $1$\\
   &   & $10$ &    &    $4.50\cdot 10^6$ & \multicolumn{1}{c|}{$405.06$} & 1.0003\\\hline
\end{tabular}
\caption{Benchmark results for our algorithm, for several parameters sets. Timings have been measured on a laptop with a 13th Gen Intel(R)
Core(TM) i7-1355U, with 16 GB RAM and running the Ubuntu 22.04.5 LTS
operating system}
\label{tab:results_final}
\end{table*}

\subsection{Benchmarks}

In this section we report about simulations for Algorithm \ref{alg:sample}.

First, we have used our implementation to sample dual-containing codes with several parameters.
For each parameters set, we have executed our algorithm 10 times, measuring both the average execution time as well as the average number of ISD calls to find a codeword of the desired weight (averaged over all runs, as well as over all steps of the algorithm).
As ISD, we have used $\leb$ with $p = 3$.
The results are shown in Table \ref{tab:results_final}.
For each parameter set, se hWe report also 

We remark that, for all the considered parameters sets, codes with the desired parameters haven always been successfully generated.
For all parameters, the average number of ISD calls is either exactly equal to 1 or only slightly larger than 1.
This means that, in the overwhelming majority of experiments (and in the overwhelming majority of steps of the algorithm), ISD has required a unique call to find a codeword of the desired weight.
This is a strong confirmation of the efficiency of our algorithm (remember the discussion in Section \ref{subsec:isd_polynomial}).

We observe that the average time increases as $v$ gets larger. This is due to the fact that, as $v$ gets higher, the value of $m_v$ gets smaller. Since, for each $\code_u$, there are on average less codewords of weight $v$, even when one ISD iteration is enough, we need to test more weight-$p$ vectors before a valid codeword is found (hence, timings increase).
Moreover, it may even happen that we need more than one ISD calls.

We observe that, when working with smaller values of $m_v^{(r)}$, the probability that our algorithm fails gets higher.
To this end, we made experiments with $n = 150$, $r = 70$ and $v = 10$.
For these parameters, $m_v^{(r)} = 1.29$.
We set the algorithm in order to halt whenever, for a value of $u$, the number of ISD calls reached 100. 
We ran the algorithm for 100 times and observed that, in 16 calls, it actually halted before completing the final step.
In particular, in 5 of these failures, the algorithm halted during step $u = r-2$ (so, with a matrix $\6H$ with 68 rows), while in the other 11 failures it halted during step $u = r-1$ (so, with a matrix $\6H$ with 69 rows). 

\subsection{Weight distributions for columns}

\begin{figure}[h!]
\resizebox{\columnwidth}{!}{
\begin{tikzpicture}
    \begin{axis}[
        xlabel={Column weight $(z)$},
        ylabel={Avg. number of weight-$z$ columns},
        xmin=1, xmax=12,
        ymin=0.001, ymax=100,
        ytick= {0.001, 0.01, 0.1, 1, 10, 100},
        legend pos = south west,
        legend style={font=\scriptsize},
        xmajorgrids=true,
        ymajorgrids=true,
        height  = 5.5cm,
        major grid style = {gray!20!white},
        width = \columnwidth,
        legend columns=1,
        ymode = log
    ]

\addplot [RoyalBlue, thick] coordinates{
(0, 5.59649833251954)
(1, 18.9177408423196)
(2, 31.4407523858269)
(3, 34.2453265423091)
(4, 27.4927269424172)
(5, 17.3475234791590)
(6, 8.95881494228869)
(7, 3.89356947190012)
(8, 1.45323367613173)
(9, 0.473040069851957)
(10, 0.135915738379999)
(12, 0.00800691442970063)
};
\addlegendentry{$n = 150$, $r = 60$, $v = 8$};

\addplot [Bittersweet, thick] coordinates{
(0, 6.17665596279054)
(1, 17.7414586165260)
(2, 24.9135376317174)
(3, 22.7932365566776)
(4, 15.2763181177733)
(5, 7.99568990845155)
(6, 3.40242123763896)
(7, 1.20997958906917)
(8, 0.366855513707143)
(9, 0.0962670496962006)
};
\addlegendentry{$n = 100$, $r = 45$, $v = 6$};

\addplot [RoyalBlue, thick, only marks, mark = x] coordinates{
(0, 0.840000000000000)
(1, 13.6500000000000)
(2, 36.4000000000000)
(3, 41.4800000000000)
(4, 31.8800000000000)
(5, 16.4400000000000)
(6, 6.62000000000000)
(7, 2.06000000000000)
(8, 0.490000000000000)
(9, 0.0900000000000000)
(10, 0.0400000000000000)
(12, 0.0100000000000000)
};

\addplot [Bittersweet, thick, only marks, mark = x] coordinates{
(0, 0.930000000000000)
(1, 15.3500000000000)
(2, 30.2900000000000)
(3, 29.4800000000000)
(4, 16.7700000000000)
(5, 5.08000000000000)
(6, 1.65000000000000)
(7, 0.360000000000000)
(8, 0.0800000000000000)
(9, 0.0100000000000000)
};

\end{axis}

\end{tikzpicture}
}
\caption{Average number of columns with weight $z$, for several code parameters. Continuous lines refer to theoretical estimates, markers refer to experimental values.
For each parameters set, we have sampled 100 codes.}
\label{fig:columns}
\end{figure}
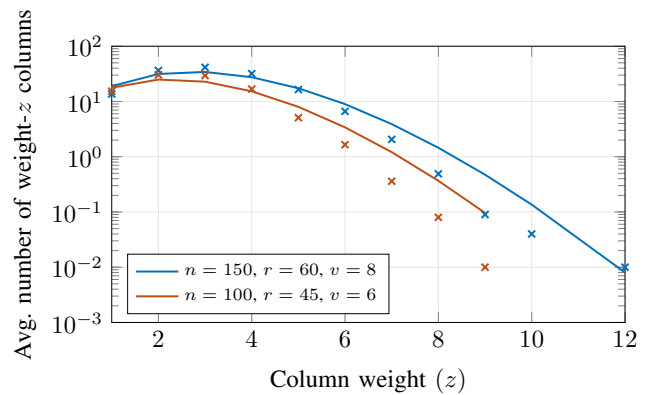

To confirm the validity of the analysis in Section \ref{subsec:unlucky}, we show that \eqref{eq:binomial_columns} yields a good estimate for the distribution of the column weight in matrices output by Algorithm \ref{alg:sample}.
A comparison between theoretical and empirical values, for some example parameter sets, is shown in Figure \ref{fig:columns}.

As we can see, the behavior of the measured distribution is well modeled by the \eqref{eq:binomial_columns}.
Interestingly, the number of null columns appears to be smaller than the theoretical estimate while, for weight $z\geq 1$, the theoretical estimate is somewhat lower.

\pagebreak

\section{Generalizations}
\label{sec:generalizations}

We briefly discuss how Algorithm \ref{alg:sample} can be generalized.
First, as we have already said, one can modify the distribution of the row weight of $\6H$ (for instance, require that the row weight follows a binomial distribution).
As soon as rows of $\6H$ are sampled independently, then the whole analysis in the paper applies; the only very little nuisance is in modifying how $\rho_w$ is computed. 
Since this would just be an exercise, we will not speak further about this generalization and, instead, we will discuss about more interesting ways to play with the algorithm. 
In particular, we will see how one can sample dual-containing LDPC codes defined over any finite field and more general general quantum LDPC codes, even those that do not follow the CSS construction.

\subsection{Dual-containing LDPC codes over non binary finite fields}

The algorithm can be easily tweaked to work any finite field $\mathbb F_q$ with $q$ elements.
ISD needs to be modified accordingly, as now enumeration of low weight vectors shall takes place over $\mathbb F_q$.
Observe that, when working over a non binary finite field, the value of $v$ does not need to be even.
With some effort, the analysis from Section \ref{sec:distributions} can be easily generalized to a $q$-ary field. 

The only meaningful difference is in how the matrices $\6H_u$ are constructed.
Indeed, remember that we were constructing $\6H_u$ by stacking all found vectors $\6h_1,\hsp\cdots,\hsp \6h_{u-1}$, together with the all ones vector which assures self-orthogonality.
The construction can remain the same only if the characteristic of the field is 2, i.e., only if $q = 2^\ell$ for some
$\ell\in\mathbb N$.
Indeed, let $\6c = (c_1,\hsp\cdots,\hsp c_n)\in\mathbb F_q^n$.
Then, self-orthogonality implies
$\sum_{i = 1}^n c^2_i = 0$.
When the characteristic is 2, we have
$$0 = \sum_{i = 1}^n c^2_i = \left(\sum_{i = 1}^n c_i\right)^2\quad \implies \quad \sum_{i = 1}^n c_i = 0.$$
Thus, in this case, the construction of $\6H_u$ can be preserved as, again, self-orthogonality can be expressed as a linear relation.

If the characteristic is different than 2, instead, this is not true anymore.
Here, the idea is to construct $\6H_u$ by stacking only $\6h_1,\hsp\cdots,\hsp \6h_{u-1}$.
Now, we cannot guarantee anymore that any low weight codeword from $\code_u$ is self-orthogonal.
In such a case, one can simply do rejection sampling and continue calling ISD until a self-orthogonal codeword is found.
Since a random $q$-ary vector is self-orthogonal with approximate probability $1/q$, on average we have to sample $q$ codewords for each step.

\subsection{Non dual-containing LDPC codes for the CSS construction}

Remember that, for the general CCS construction, we need two sparse, full rank matrices $\6H_1\in\mathbb F_2^{r_1\times n}$ and $\6H_2\in\mathbb F_2^{r_2\times n}$ such that
\begin{equation}
\6H_1\6H_2^\top = \60.
\end{equation}
Let $\code_1$ and $\code_2$ be the codes whose parity-check matrices are $\6H_1$ and $\6H_2$, respectively.
Since codewords of $\code_1^\bot$ are orthogonal to $\6H_2$, they are contained in $\code_2$.
We thus have $\code_1^\bot \subseteq \code_2$.

Sampling such codes is rather simple.
Assume, for simplicity, that we want rows of $\6H_1$ and $\6H_2$ with weights $w$ and $v$, respectively.
Then, we can sample $\6H_2$ from $\mathcal H_{n,r,v}$.
If the value of $m_w$ is greater than $1$, then with high probability $\code_2$ contains codewords of weight $w$.
In particular, we should have $m_w \geq r_1$, since for $\6H_1$ we need $r_1$ rows.
These rows can be sampled from $\code_2$ efficiently with ISD. 

\subsection{Stabilizer LDPC codes}

We now tackle the more general case of a stabilizer code that does not necessarily follow the CSS construction.
Remember that, in this case, we need to sample two matrices $\6H_X, \6H_Z\in\mathbb F_2^{r\times n}$ such that 
$$\6H_X\6H_Z^\top + \6H_Z\6H_X^\top = \60.$$
Let us express such matrices as
$$\6H_X = \begin{pmatrix}\6a_1\\\vdots\\ \6a_r\end{pmatrix},\quad \6H_Z = \begin{pmatrix}\6b_1\\\vdots\\ \6b_r\end{pmatrix}.$$
The symplectic product condition results into
\begin{equation}
\label{eq:new_sympl}
\6a_i\6b_j^\top+\6b_i\6a_j^\top = 0,\quad \forall i,j.
\end{equation}
Notice that this is trivially satisfied whenever $i = j$, so we just need to consider the case $i\neq j$.

Let us assume we already have the first row for both $\6H_X$ and $\6H_Z$ and want to sample the second row.
In this case, we just need to satisfy \eqref{eq:new_sympl} for $i = 1$ and $j = 2$, which results in
$$\6a_1\6b_2^\top+\6b_1\6a_2^\top = 0 \quad \implies \quad (\6b_2,\hsp\6a_2)\in\mathrm{Ker}\big((\6a_1,\hsp \6b_1)\big).$$
When it comes to the third row, we have the following new constraints which must be satisfied:
$$\begin{cases}
\6a_1\6b_3^\top+\6b_1\6a_3^\top = 0,\\
\6a_2\6b_3^\top+\6b_2\6a_3^\top = 0.
\end{cases}
$$
Notice that this implies $(\6b_3,\hsp\6a_3)\in\mathrm{Ker}\big((\6a_1,\hsp \6b_1),\hsp (\6a_2,\hsp \6b_2)\big)$.

Generally, let us assume we have generated the first $u-1$ rows of $\6H_X$ and $\6H_Z$ and want the $u$-th row for both matrices.
These rows must satisfy
$$(\6b_u,\hsp\6a_u)\in\mathrm{Ker}\big((\6a_1,\hsp \6b_1),\hsp (\6a_2,\hsp \6b_2),\hsp \cdots,\hsp (\6a_{u-1},\hsp \6b_{u-1})\big).$$
To keep the notation consistent with that of Algorithm \ref{alg:sample}, let 
$$\6H_u = 
\begin{pmatrix}
\6a_1 & \6b_1 \\
\6a_2 & \6b_2 & \\
\vdots & \vdots \\
\6a_{u-1} & \6b_{u-1}
\end{pmatrix}\in\mathbb F_2^{(u-1)\times 2n},$$
and $\code_u\subseteq \mathbb F_2^{2n}$ be the code whose parity-check matrix is $\6H_u$.
Then, we can efficiently search for $(\6b_u,\hsp \6a_u)\in\code_u$ using ISD and, whenever we find it, we can proceed with the subsequent value of $u$.

Observe that there is no meaningful difference with respect to the cases we have already treated.
Indeed, we just need to choose parameters so that, at every step, the desired EWD coefficient is high enough.

\section{Conclusions}
In this paper we have presented and analyzed an algorithm to sample random quantum LDPC codes.
Technically, this boils down to sampling sparse matrices that satisfy certain orthogonality relations.
Our analysis shows that we can efficiently sample codes with code length $n$ and constant code rate $R$, with parity-check matrices where the row weight is $O\big(\log_2(n)\big)$.
Numerical simulations confirm the efficiency and feasibility of our algorithm.

As we have discussed in the paper, there are many open questions as well as possibilities for follow-up works.
In particular, we believe focusing on further generalizations of the algorithm (e.g., in order to avoid short cycles as much as possible) represents an interesting research direction.
Analogously, the study of an ISD algorithm which is specifically tailored for LDPC codes, apart from being a research direction of independent interest, should improve significantly the efficiency of our algorithm.

While we have not been able to formally prove that our algorithm runs in expected polynomial time, there are strong hints that this indeed happens for many ranges of parameters.
We see this as another relevant open question and we plan to focus on this task in follow up works.

\section*{Acknowledgments}
The author wishes to thank Alessio Baldelli and Michele Pacenti for being so kind and helpful in clarifying many doubts about  quantum codes.

This work was partially supported by project SERICS (PE00000014) under the Italian Ministry of University and Research (MUR) National Recovery and Resilience Plan, funded by the European Union - Next Generation EU.,

\printbibliography

\end{document}